# Inter-Block Permuted Turbo Codes[†]

Yan-Xiu Zheng and Yu T. Su

Department of Communications Engineering

National Chiao Tung University

1001 Dar-Shei Rd., Hsinchu, 30056, TAIWAN

Tel.: +886-3-571-2121,   Fax: +886-3-571-0116

E-mail: ytsu@cc.nctu.edu.tw

**Abstract**

The structure and size of the interleaver used in a turbo code critically affect the distance spectrum and the covariance property of a component decoder's information input and soft output. This paper introduces a new class of interleavers, the inter-block permutation (IBP) interleavers, that can be build on any existing "good" block-wise interleaver by simply adding an IBP stage. The IBP interleavers reduce the above-mentioned correlation and increase the effective interleaving size. The increased effective interleaving size improves the distance spectrum while the reduced covariance enhances the iterative decoder's performance. Moreover, the structure of the IBP(-interleaved) turbo codes (IBPTC) is naturally fit for high rate applications that necessitate parallel decoding.

We present some useful bounds and constraints associated with the IBPTC that can be used as design guidelines. The corresponding codeword weight upper bounds for weight-2 and weight-4 input sequences are derived. Based on some of the design guidelines, we propose a simple IBP algorithm and show that the associated IBPTC yields 0.3 to 1.2 dB performance gain, or equivalently, an IBPTC renders the same performance with a much reduced interleaving delay. The EXIT and covariance behaviors provide another numerical proof of the superiority of the proposed IBPTC.

† This work is supported by the National Science Council of Taiwan under Contract 92-2213-E-009-050. Parts of this paper were presented at the IEEE Vehicular Technology Conference, Vancouver, Canada, September 2002 and at the International Symposium on Turbo Codes, Brest, France, September 2003.

# I. Introduction

The turbo code's extraordinary performance is in part due to a class of suboptimal iterative decoding algorithms that generate soft outputs based on the maximum a posteriori (MAP) principle. At each decoding round, an a posteriori probability (APP) decoder provides extrinsic information for use in the ensuing round as the a priori information. The extrinsic information about an information bit is gathered through message-passing from the channel output samples corresponding to other related bits, and its derivation is based on the structures of the interleaver and the component codes as well as the statistical property of the channel. The interleaver is thus an integrated and critical component of a turbo code (TC) and its importance has been well documented.

As the interleaver is of finite size, say, $L$ bits, the input information sequence is segmented into blocks of $L$ bits, leading to separate encoding, interleaving and decoding of each block. As a result, the message-passing process induced by the interleaver is confined to within a block. Given the component codes and the decoding algorithm, performance can be improved by increasing the interleaver (block) size.

At the cost of prolonged interleaver delay, the increased block size not only enables the decoder to gather the extrinsic information from more (noisy) data and parity samples but also makes it easier to reduce the covariance between the extrinsic (information) outputs and the corresponding information inputs, which, as suggested by Hokfelt *et al.* [7] and Sadjadpour *et al.* [8], is a desired attribute of the interleaver/deinterleaver pair. Intuitively the less covariance between information input and extrinsic output the more 'new' information the extrinsic output would carry over to the new APP decoding round, where two consecutive decoding rounds constitute a decoding iteration. Increasing the interleaver size also provides greater flexibility in designing a better permutation rule to avoid generating low weight codewords and to improve the free distance and spectrum properties of the associated turbo code [19], [20].

In short, the interleaver and the component codes' structures together determine the distance spectrum of the code and, through the message-passing process embedded in the iterative decoding procedure, determine the bit error performance. In this paper, we present a new interleaver structure that serves both purposes without increasing the total interleaving delay if certain degrees of parallelism are allowed and a proper decoding schedule is in place.

Before formally defining our interleaver structure, we give a tutorial explanation for its effectiveness in expanding an iterative decoder's message-passing range. Consider the interleaver structure shown in Fig. 1. A data sequence is partitioned into $L$-bit blocks, each block being represented by a rectangular. Information bits are first permuted within their respective blocks. After such intra-block permutations, bits within a block are further systematically permuted with those in other neighboring blocks. It is obvious that the additional inter-block permutation (IBP) leads to progressively larger ranges for message-passing as the number of decoding iteration increases. Intuitively, when the extrinsic information about a particular information bit is gathered from more and farther data and parity samples, it becomes more reliable and less correlated with the information bit. Let us elaborate on this message-passing process associated with the iterative decoding of an IBP-interleaved turbo code (IBPTC).

In general, a turbo code may consist of several parallel constituent codes and interleavers. For simplicity of presentation, however, we will consider only the classic turbo codes that use two identical recursive systematic convolutional component codes and one interleaver in the subsequent discussion. Fig. 2 is a graph representation that illustrates the behavior of such a classic turbo code that uses an IBP interleaver (IBPI). In this figure, nodes $U_i, U_i'$ represent a data block of length $L$ and its permuted version. $X_i^0, X_i^1, X_i^2$ denote the uncoded (systematic part) and two encoded blocks while $Y_i^0, Y_i^1, Y_i^2$ are the corresponding received blocks. The bold solid lines connecting $U_i$ and $U_i'$ indicate intra-block permutations that confine permuted bits to within the same block. An IBP that permutes bits of a given block with some of those bits within the two immediate adjacent blocks and the original block is represented by the dotted lines connecting $U_i$ and $\{U_{i+1}, U_i, U_{i-1}\}$. Both solid and dotted lines also represent directions of information flow to and from a node during an iterative decoding process.

Take node $U_3$ as an example and, for the convenience of describing the decoding process, refer to the decoder responsible for decoding node $U_i$ or the corresponding codeword $(X_i^0, X_i^1)^1$ as the first APP decoder and that in charge of decoding $U_i'$ or its associated codeword the second APP decoder–although physically they might be the same one. After the first APP decoding of node $U_3$, the associated extrinsic information is passed to nodes $U_2', U_3'$ and $U_4'$ by the interleaver. When the second APP decoder is decoding the $U_2'$ ($U_4'$) block, it uses extrinsic information derived from $U_3$, along with that derived from $U_1$ and $U_2$ ($U_4, U_5$), as the a priori information to generate new extrinsic information, after de-interleaving, to nodes $\{U_1, U_2, U_3\}$ ($\{U_3, U_4, U_5\}$). We see that, after just one iteration the information associated with the $U_3$ block has already been passed to portions of four neighboring blocks and $U_3$ itself. Therefore, when the first APP decoder is decoding node $U_3$ for the second time it can use information collected from these four adjacent blocks. As one proceeds with further iterations, more information will be available for the APP decoder while, as we shall see in the next section, the interleaving delay can be kept at a fixed $2L$ bits, and, more importantly, the average decoding delay per block can remain invariant if certain conditions are satisfied.

---
[1]For brevity, we shall make no distinction between $U_i$ (or $U_i'$) and its associated codeword if there is no danger of confusion.

This last property is critically important, for it implies that an IBP interleaver can have an unbounded equivalent interleaving depth (size) that is constrained only by the number of turbo decoding iterations and the data sequence duration while keeping the average interleaving and decoding delay bounded by its local interleaving depth.

The rest of this paper is organized as follows: In Section II, we introduce the structure of IBP interleavers and define the related parameters. In section III, latency and implementation issues are discussed. Some important codeword weight bounds and intra-block interleaver constraints associated with the IBPTC are derived in Section IV. When deriving these bounds, we first assume no knowledge of the intra-block permutation then obtain some constraints on the intra-block permutation to avoid producing low-weight codewords. In Section V, we consider the problem of joint intra-block and inter-block permutation design and derive an upper-bound for the achievable minimum distance. Some of the properties derived in Sections IV and V can be used as the IBPI design guidelines. Architecture for realizing IBPTC and an IBP algorithm are presented in Section VI, so is a modified semi-random interleaver which serves as the intra-block interleaver. In Section VII, we provide simulation results to validate our assertion that our proposal does yield significant improvement. The final section summarizes our work and gives some concluding remarks.

## II. INTER-BLOCK PERMUTATION INTERLEAVER

### A. Interleaver description

Let $\mathbf{u} = \{u_k\}_{k=-\infty}^{k=\infty}$ be an input data sequence and $\pi$ be the interleaver that maps $\mathbf{u}$ into $\mathbf{u}' = \{u'_k\}_{k=-\infty}^{k=\infty}$ such that $u'_{\pi(k)} = u_k$, where $\pi(k)$ is the permutation rule corresponding to $\pi$. $\pi^{-1}$ denotes the permutation of the de-interleaver of $\pi$, representing the inverse mapping that maps $\mathbf{u}'$ back to $\mathbf{u}$, i.e., $u_{\pi^{-1}(k)} = u'_k$, where $\pi^{-1}(k)$ is the inverse permutation rule corresponding to $\pi^{-1}$. Denote the block interleaver by $\pi_{block}$ and the corresponding permutation rule by $\pi_{block}(k)$, $0 \leq k < L$, $L$ being the length of the block interleaver. Using the abbreviated notations $||k||_L \overset{def}{=} k \bmod L$ and $|k|_L \overset{def}{=} \lfloor k/L \rfloor$ where $\lfloor x \rfloor$ is the integer part of $x$, we denote the intra-block interleaver and the inter-block interleaver by $\pi_{intra}$ and $\pi_{inter}$ and define the corresponding permutation rules by

$$\pi_{intra}(k) = L|k|_L + \pi_{block}(||k||_L) \tag{1}$$

$$\pi_{inter}(k) = L\left(|k|_L + f_{in}(k)\right) + f_{ib}(k). \tag{2}$$

Similarly, the inverse intra-block and inter-block permutation rules are defined by

$$\pi_{intra}^{-1}(k) = L|k|_L + \pi_{block}^{-1}(||k||_L) \tag{3}$$

$$\pi_{inter}^{-1}(k) = L\left(|k|_L + f_{in}^d(k)\right) + f_{ib}^d(k), \tag{4}$$

The IBP rule is characterized by two functions, $f_{ib}(k)$ and $f_{in}(k)$ and its inverse is characterized by $f_{ib}^d(k)$ and $f_{in}^d(k)$; all four are integer-valued functions. $0 \leq f_{ib}(k), f_{ib}^d(k) < L$ represent the relative positions within a block after the inter-block interleaving and deinterleaving. $f_{in}(k)$ and $f_{in}^d(k)$ are block indicator functions that determine to and from which block the $k$th bit is moved by these two operations. $f_{in}(k) = 0$ or $f_{in}^d(k) = 0$ means that the $k$th symbol remains in the same block and the range constraints, $-S_b \leq f_{in}(k) \leq S_f$ and $-S_f \leq f_{in}^d(k) \leq S_b$, define the forward and backward interleaver spans $S_f$ and $S_b$.

An IBPI is then defined by $\pi_{ibp}(k) = \pi_{inter}(\pi_{intra}(k))$, the composition of the intra- and inter-block permutation rules, while the IBP de-interleaver (IBPDI) is the composition of the corresponding inverse rules, $\pi_{ibp}^{-1}(k) = \pi_{intra}^{-1}(\pi_{inter}^{-1}(k))$. They can be expressed as

$$\pi_{ibp}(k) = \hat{f}_{in}(k)L + \hat{f}_{ib}(k), \tag{5}$$

$$\pi_{ibp}^{-1}(k) = \hat{f}_{in}^d(k)L + \hat{f}_{ib}^d(k), \tag{6}$$

where

$$\hat{f}_{in}(k) = |k|_L + f_{in}(\pi_{intra}(k)), \tag{7}$$

$$\hat{f}_{ib}(k) = f_{ib}(\pi_{intra}(k)), \tag{8}$$

$$\hat{f}_{in}^d(k) = |k|_L + f_{in}^d(k), \tag{9}$$

$$\hat{f}_{ib}^d(k) = \pi_{block}^{-1}\left(f_{ib}^d(k)\right). \tag{10}$$

*B. Interleaving and deinterleaving delays*

Define the interleaver delay, $D_i$, and the deinterleaver delay, $D_d$, by

$$D_i = \max_k\{k - \pi(k)\}, \quad D_d = \max_k\{k - \pi^{-1}(k)\}. \tag{11}$$

The interleaving and deinterleaving delays per decoding round of an IBPI are bounded by

$$
\begin{aligned}
D_{i,IBP} &= \max_k \{k - \pi_{ibp}(k)\} = \max_k \left\{ k - \hat{f}_{in}(k)L - \hat{f}_{ib}(k) \right\} \\
&= \max_k \left\{ k - (|k|_L - S_b)L - \hat{f}_{ib}(k) \right\} \leq (S_b + 1)L
\end{aligned}
\tag{12}
$$

and

$$
\begin{aligned}
D_{d,IBP} &= \max_k \{k - \pi_{ibp}^{-1}(k)\} = \max_k \left\{ k - \hat{f}_{in}^d(k)L - \hat{f}_{ib}^d(k) \right\} \\
&= \max_k \left\{ k - (|k|_L - S_f)L - \hat{f}_{ib}(k) \right\} \leq (S_f + 1)L.
\end{aligned}
\tag{13}
$$

A fully dispersed interleaver (deinterleaver) is one that achieves the corresponding upperbound, and a symmetric interleaver (deinterleaver) is one with the same forward and backward spans, $S_f = S_b = S$.

### C. Special inter-block interleavers

Five IBPIs deserve special attention.

*Definition 1:* If $f_{ib}(k) = ||k||_L, \forall\ k$, then the corresponding inter-block interleaver is called a *Type I (inter-block) interleaver*.

*Definition 2:* An inter-block interleaver is a *Type II (inter-block) interleaver* if $f_{in}(k) = f_{in}(k + nT_s)$, $\forall\ n$, $L|k|_L \leq k, k + nT_s < L(|k|_L + 1)$, where $T_s = S_f + S_b + 1$, and the integer-valued function $f_{in}(k)$ is injective within a period.

*Definition 3:* An inter-block interleaver is a *Type III (inter-block) interleaver* if $f_{in}^d(k) = f_{in}^d(k + nT_s)$, $\forall\ n$ and $L|k|_L \leq k + nT_s < L(|k|_L + 1)$, where $T_s = S_f + S_b + 1$, and the integer-valued function $f_{in}^d(k)$ is injective within a period.

*Definition 4:* An inter-block interleaver that possesses all the properties of the *Types I, II, III* (inter-block) interleavers is a *Type IV (inter-block) interleaver.*

*Definition 5:* An interleaver is a *swap interleaver* if $\forall\ i$, $\pi(i) = j \Rightarrow \pi(j) = i$, i.e. $\forall\ i, \pi(i) = \pi^{-1}(i)$.

Each category of interleavers has some desired properties (e.g., *Types II* and *III* interleavers are locally or blockwise periodic) that will be proved useful in our interleaver design. For *Type I* interleaver, $\hat{f}_{ib}(k) = \pi_{block}(||k||_L)$, i.e., the relative coordinate of a symbol within a block is invariant to the inter-block permutation. *Definition 2* implies that *Type II* interleavers are periodic and $||f_{in}(\pi_{intra}(x)) - f_{in}(\pi_{intra}(y))||_{T_s} = ||\pi_{block}(||x||_L) - \pi_{block}(||y||_L)||_{T_s}$. We also note that the *swap interleavers* have a simple symmetric structure that requires less storage size for implementation.

## III. Latency and implementation concerns

Although the interleaving process of an IBPI is defined by the composition of the intra- and inter-block permutations, it can be implemented by a single step. The encoder knows to which position each bit (or sample) in a given block should be moved and can do so immediately after it receives each incoming bit. But to encode a given, say the $i$th, interleaved block $U_i'$ into $X_i^2$, it has to wait until the complete $(i+S)$th block is received. In a sense, IBP is a non-causal operator. The time elapsed between the instant the encoder receives the first bit of the $i$th block and the moment when it receives the last bit of the $(i+S)$th block and outputs its first encoded bit of $X_i^2$ is simply $(1+S)L$-bit durations. By contrast, a classic TC with a block size of $L$ bits has an encoding delay of approximately $L$ bits.

The interleaving (or deinterleaving) delay per decoding round, or the single-round interleaving delay (SRID) for short, is proportional to the encoding delay. But the decoding delay of an IBPTC decoder is a much more complicated issue. For the first decoding of each received block, there can be zero waiting time, but for later decoding rounds the corresponding delays depend on, among other things, the decoding schedule used. With the same block size, the decoding delay of the first received block for the classic TC is definitely shorter than that for the IBPTC. But if one considers a period that consists of multiple blocks (otherwise one will not have enough blocks to perform inter-block permutation) and takes the decoding schedule into account, then the average decoding latency difference can be completely eliminated. This is because the APP decoder (including the interleaver and deinterleaver) will not stay idle until all blocks within the span of a given block are received. Instead, the APP decoder will perform decoding-interleaving or deinterleaving operations for other blocks according to a predetermined decoding schedule before it can do so for the given block (and the given decoding round).

If we define the total decoding delay (TDD) as the time span between the instant a decoder receives the first input sample (from the input buffer) and the moment when it outputs its last decision then both the IBP and classic approaches yield the same TDD even if only one APP decoder is used. We use the following example and Fig. 3 to support our claim; its generalization is straightforward.

Suppose we receive a total of 7 blocks of samples (in a packet, say) and want to finish decoding in 2 iterations (4 decoding rounds). One can easily see from Fig. 3 that a classic TC decoder would output the first decoded block in 4 $DT$ cycles, where $DT$ is the number of cycles needed to perform a single-block APP decoding plus SRID, while the IBPTC decoder needs 10 DT cycles to output its first decoded block. However, if one further examines the decoding delays associated with the remaining blocks, then one finds they are 8, 12, 16, 20, 24, and 28 DT cycles for the classic TC decoder while those for the IBPTC decoder are 14, 18, 22, 25, 27 and 28 DT cycles, respectively. So in the end, both approaches reach the final decision at the same time.

In general, except for the first block and the last $2N-1$ blocks, both decoders result in a constant delay of $2N$ DT cycles between two adjacent output blocks, $N$ being the number of decoding iterations. The IBPTC decoder with $S=1$ requires a first-block decoding delay (FBDD) of $N(1+2N)$ DT cycles while the FBDD for the classic TC is only $2N$ DT cycles. The inter-block decoding delays (IBDD, i.e., decoding latency between two consecutive output blocks) for the last $2N-1$ output blocks of an IBPTC decoder using a decoding schedule similar to that shown in Fig. 3 form a monotonic decreasing arithmetic sequence $\{2N-1, 2N-2, \cdots, 1\}$ (in DT cycles). The IBDD of a classic TC decoder remains a constant $2N$ DT cycles. On the average, both codes give the same IBDD.

The above assessment on the decoding delay is made under the assumption that both codes use the same block size $L$ and a single APP decoder is used. As was mentioned in Section I, an IBPTC has an equivalent interleaving depth that grows as the number of iterations increases. We will show later that with an identical block size an IBPTC always outperforms its classic counterpart. In other words, an IBPTC requires a smaller block size and thus less decoding delay to achieve the same performance.

We now suggest an alternative viewpoint on the concepts of IBP based on the the above example. For a classic TC with block size of $7L$ bits, the FBDD is 28 DT cycles. But if one divides a $7L$-block into 7 subblocks and a special interleaver which performs successive intra-subblock and inter-block permutations on these subblocks, the corresponding decoding delays in DT cycles for these subblocks are 14, 18, 22, 25, 27, and 28, respectively. Therefore, although both code structures result in identical TDD the IBPTC structure is able to supply partial decoded outputs much earlier.

Since the classic TC can use a better interleaver of depth $7L$ and provide better performance, the class of IBPTCs offers an alternative tradeoff between decoding delay and performance. The relative performance degradation can at least be partially recovered by increasing the number of blocks (or subblocks) involved in decoding and the decoding latency can be further reduced by employing multiple APP decoders; see Fig. 4 for a typical decoding schedule using 4 APP decoders. The regularity feature of an IBPI makes IBPTCs naturally suited for parallel decoding–which is often a necessity in high data rate applications. By contrast, parallel decoding of a classic TC is very likely to have difficulty in finding a memory efficient solution for the memory bank access collision (i.e., more than one APP decoder try to access the same information simultaneously) problem.

Fig. 5 (a) shows an IBPTC decoding module for one iteration. The parenthesized numbers in each block indicate the corresponding latencies. A pipeline structure similar to that proposed by Hall [3] is shown in Fig. 5(b). The pipeline structure renders short decoding latency at the expense of increased complexity that is proportional to the number of the decoding modules.

But the pipeline decoder architecture is not necessarily the optimal solution in terms of performance and average decoding latency. It has been shown [9] that with an appropriate early-termination scheme, multiple, say $M$ ($2 \leq M <$ number of decoding rounds) APP decoders and proper scheduling, one can improve the BER performance of an IBPTC and reduce the associated average iteration number. Although early-termination schemes can also be applied to classic TCs, not much performance gain is expected and the reduction of the average decoding iteration number is limited since early-terminated blocks are unable to pass the highly reliable extrinsic information output to un-terminated blocks.

[9] further shows that the structure of the IBPTC renders its decoding amenable for highly dynamic decoding schedules that are both distributive and cooperative: sharing all modularized decoding resources–the APP component decoders, interleavers/deinterleavers, memory–while passing information amongst component decoders. Depending on the decoding schedule and the degree of parallelism, the IBPTC admits a variety of decoder architectures that meet a large range of throughput and performance demands. Its performance can be improved by using a proper decoding schedule, increasing the block size, the interleaving span, the number of decoding iterations and the number of blocks involved in decoding. The availability of these options are indications of the flexibility and versatileness of the IBPTC.

We also want to remark that our proposal is also an attractive option for applications like deep space communication systems in which the earth decoder can enhance its BER performance by increasing the number of iterations (thereby increasing the equivalent interleaving size and the FBDD) while the encoder in the space segment remains intact.

The issues of IBPTC decoder architecture, the associated decoding schedule and memory management, though interesting and worthy of detailed investigation, are beyond the scope of this paper and will not be discussed henceforth.

## IV. IBP PROPERTIES

### A. Basic definitions and results

Consider a turbo code $\mathbf{C}$ that consists of $m$ rate $1/2$ recursive systematic component codes. Denote by $\mathbf{u} = \{\cdots, u_{-1}, u_0, u_1, u_2, \cdots\}$ a binary input sequence, and by $\mathbf{X} = \{\mathbf{x}^0, \mathbf{x}^1, \cdots, \mathbf{x}^m\}$ the codeword associated with $\mathbf{u}$, where $\mathbf{x}^i$ is the output parity-bit sequence of the $i$th component code while $\mathbf{x}^0 = \mathbf{u}$ represents both the input sequence and the systematic (uncoded) output sequences. A minimum weight codeword thus consists of the sequences $\{\mathbf{x}_{min}^i, i = 0, 1, \cdots, m\}$. For a bi-infinite sequence $\mathbf{x}^i = \{\cdots, x_{-2}^i, x_{-1}^i, x_0^i, x_1^i, x_2^i, \cdots\}$, the sequence $\mathbf{x}_k^i = \{\cdots, 0, 0, x_{kL}^i, x_{kL+1}^i, \cdots, x_{(k+1)L-1}^i, 0, \cdots\}$ is called the $k$th *block-matched* (BM) sequence of $\mathbf{x}^i$. We consider the special case $m = 2$ only and refer to the corresponding turbo code using a conventional block-wise interleaver as a classic TC $\mathbf{C}_b$.

Depending on the hypertrellis connections between the adjacent blocks, we can use one of the three encoding/decoding options, namely continuous, truncated and terminated 'co-decoding' [6]. In accordance with these options, we define a *continuous* IBPTC (C-IBPTC) as one that encodes each data block using the end state of the previous coded block as the initial state and adds the tail-bits only for the last data block. On the other hand, a *discontinuous* IBPTC (D-IBPTC) encodes each data block individually, either by appending the tail-bits at the end of a block or by using the tail-biting encoding. We refer to the former class as the tail-padding IBPTC (TP-IBPTC) while the latter class as the tail-biting IBPTC (TB-IBPTC).

The following theorem specifies the conditions under which the free distance of a C-IBPTC will be greater than or at least equal to that of its corresponding classic TC.

*Theorem 1:* For a classic TC $\mathbf{C}_b$, the corresponding C-IBPTC $\mathbf{C}_{ibp}$ based on (5)-(10) has a free distance greater than or equal to that of $\mathbf{C}_b$ if a *Type I* inter-block interleaver is used and all BM sequence pairs of a minimum weight codeword of $\mathbf{C}_{ibp}$, $\{\mathbf{x}_{min}^i, i = 0, 1, 2\}$, are also valid codewords of the corresponding component codes.

*Proof:* For a C-IBPTC $\mathbf{C}_{ibp}$, there exists at least a finite-weight data sequence $\mathbf{u}_{min} = \mathbf{x}_{min}^0$ whose corresponding codeword has the minimum weight. Suppose the nonzero elements of $\mathbf{u}_{min}$ are at positions $\{k_1, k_2, \ldots, k_n\}$ and the corresponding codeword is $\bar{\mathbf{X}}_{min} = \{\bar{\mathbf{x}}^0, \bar{\mathbf{x}}^1, \bar{\mathbf{x}}^2\}$. We partition $\bar{\mathbf{x}}^i$ into blocks of equal length $L$ bits and construct the associated BM sequences. The $j$th BM sequence and its IBP interleaved version generate, for the two component codes, the encoded parity-bit sequences, $\bar{\mathbf{x}}_j^1$ and $\bar{\mathbf{x}}_j^2$ with Hamming weights $w_t(\bar{\mathbf{x}}_j^1)$ and $w_t(\bar{\mathbf{x}}_j^2)$, respectively. The systematic parts of both component codes are the same and the corresponding BM sequences are denoted by $\bar{\mathbf{x}}_j^0$. If $f_{ib}(k) = ||k||_L$, taking module $L$ on $\pi_{ibp}(k_i)$ gives

$$||\pi_{ibp}(k_i)||_L = ||L\hat{f}_{in}(k_i) + \hat{f}_{ib}(k_i)||_L = ||f_{ib}(\pi_{intra}(k_i))||_L$$
$$= ||\pi_{intra}(k_i)||_L = ||L|k|_L + \pi_{block}(||k_i||_L)||_L = \pi_{block}(||k_i||_L). \tag{14}$$

Let $M_l$ be the permutation defined on the space of all binary BM sequences that moves the matched length-$L$ block of a BM sequence to the $l$th block, i.e., $M_l : \mathbf{x}_k^i \to \mathbf{x}_l^i, \forall k$. As the 2-tuple $(\bar{\mathbf{x}}_j^0, \bar{\mathbf{x}}_j^1)$ is a valid codeword of the first component code of $\mathbf{C}_{ibp}$ according to our assumption, $\left( \bigoplus_j M_l(\bar{\mathbf{x}}_j^0), \bigoplus_j M_l(\bar{\mathbf{x}}_j^1) \right)$, $\bigoplus$ denoting addition of binary vectors, is also a codeword of the same component code. (14) implies that $\left( \bigoplus_j M_l(\tilde{\mathbf{x}}_j^0), \bigoplus_j M_l(\bar{\mathbf{x}}_j^2) \right)$ are valid codewords for the second component code of $\mathbf{C}_{ibp}$ and $\mathbf{C}_b$ since the additional IBP does not change the relative positions of input bits within a block, where $\tilde{\mathbf{x}}_j^0$ is the IBP-interleaved version of $\bar{\mathbf{x}}_j^0$. The inequality

$$w_t(\mathbf{x}_j^i) + w_t(\mathbf{x}_l^i) = w_t(M_l(\mathbf{x}_j^i)) + w_t(M_l(\mathbf{x}_l^i)) \geq w_t\left( M_l(\mathbf{x}_j^i) \oplus M_l(\mathbf{x}_l^i) \right), \ \forall j \neq l \tag{15}$$

then implies that the free distance of $\mathbf{C}_{ibp}$, $d_{free}(\mathbf{C}_{ibp})$ satisfies

$$d_{free}\left(\mathbf{C}_{ibp}\right) = \sum_i \sum_j w_t(\bar{\mathbf{x}}_j^i) \geq \sum_i w_t\left( \bigoplus_j M_l(\bar{\mathbf{x}}_j^i) \right) \geq d_{free}\left(\mathbf{C}_b\right) \tag{16}$$

■

For a D-IBPTC, the "sub-codewords" associated with each input block automatically satisfy the requirement on the BM sequences. Since both tail-padding and tail-biting convolutional codes are linear codes, we have

*Corollary 1:* For a classic TC $\mathbf{C}_b$, the corresponding D-IBPTC $\mathbf{C}_{ibp}$ based on (5) has a free distance greater than or equal to that of $\mathbf{C}_b$ if $\pi_{inter}$ is a *Type I* inter-block interleaver.

Define two equivalent relations "$\sim$" and "$\cong$" on the set of integers $Z$ by

$$||i - j||_{T_c} = 0 \iff i \sim j$$

$$|i|_L = |j|_L \iff i \cong j$$

where $i, j \in Z$ and $T_c$ is the period (to be defined in the next paragraph) of a recursive systematic convolutional (RSC) code. Clearly, $i \sim j$ or $i \ncong j$ means $i$ is equivalent to $j$ in either sense.

[26] and [28] show that the encoder of an RSC code acts like a scrambler or equivalently, a linear time-invariant IIR filter, on the input sequence and can be realized by using a shift register with both feedback and feedforward branches. It is obvious that such an encoder would have a periodic impulse response. The rate $1/2$ RSC code is specified by the transfer matrix $[1, g_f(D)/g_b(D)]$, where $g_b(D)$ is usually a primitive binary polynomial of degree $m$. The period of the impulse response of the non-systematic part, $g_f(D)/g_b(D)$, is given by $T_c$ whose maximum value is $2^m - 1$. We denote by $\mathbf{u}_{ij} = \{u_k\}$ a weight-2 input sequence whose only nonzero elements are at coordinates $i$ and $j$; the corresponding codeword is denoted by $\mathbf{X}_{ij}$. Therefore, $T_c$ is also the smallest integer such that $\mathbf{u}_{ij}, i \sim j$, will generate a finite-weight output parity sequence. It is thus easy to show [26]

*Lemma 1:* Let $\mathbf{u}_{ij}$ be the input sequence to a scrambler with period $T_c$ and $scrb(\mathbf{u}_{ij})$ be the corresponding output parity sequence. If $i \sim j$, then there exists $\alpha \in \mathcal{N}$ and $\beta \in Z$ such that $w_t(scrb(\mathbf{u}_{ij})) = \alpha|i - j|/T_c + \beta$, where $\mathcal{N}$ is the set of positive integers and $\alpha, \beta$ depend on the encoder (scrambler) structure.

Obviously, if $i \nsim j$, $\mathbf{u}_{ij}$ will generate an infinite weight parity sequence if there is no termination at the end of a block. *Lemma 1* implies that the codeword weight, $w_t(\mathbf{X}_{ij})$, of a classic rate $1/3$ turbo code satisfies

$$w_t(\mathbf{X}_{ij}) \geq 2 + \alpha \left( \frac{|i - j| + |\pi(i) - \pi(j)|}{T_c} \right) + 2\beta, \tag{17}$$

with equality holds iff

$$i \sim j \text{ and } \pi(i) \sim \pi(j). \tag{18}$$

*B. Bounds on codeword weights associated with weight-2 input*

Define $\tilde{w}_{2,min} \overset{def}{=} \min_{(i,j) \in s_m} w_t(\mathbf{X}_{ij})$, where $s_m \overset{def}{=} \{(i,j)|i \sim j, \pi_{ibp}(i) \sim \pi_{ibp}(j)\}$ and let

$$\delta_{min} = \min_{(i,j) \in s_m} [|i - j| + |\pi_{ibp}(i) - \pi_{ibp}(j)|]. \tag{19}$$

For the class of C-IBPTCs, $\tilde{w}_{2,min} = w_{2,min} \overset{def}{=} \min_{(i,j)} w_t(\mathbf{X}_{ij})$, therefore, maximizing the minimum weight of the codewords associated with the weight-2 input sequences is equivalent to maximizing $\delta_{min}$. The next theorem provides an upper-bound of $w_{2,min}$ any IBPTC can achieve, if choosing the intra-block interleaver is not an option.

*Theorem 2:* For an IBPTC using an inter-block interleaver, $\pi_{inter}$, there exists an $\pi_{intra}$ such that $w_{2,min} \leq 2 + \alpha(S_f + S_b + 2) + 2\beta$, if $L > T_c(S_f + S_b + 1)$.

*Proof:* Consider the partition $\{0, 1, \cdots, L - 1\} = \bigcup_{j=0}^{T_c - 1} \bigcup_{l=-S_b}^{S_f} S_{jl}$, where $S_{jl} = \{m|m \in S_j, f_{in}^d(m) = l\}$, $S_j = \{n|n \sim j, \ 0 \leqslant n < L\}$, $0 \leq j < T_c$ and $-S_b \leq l \leq S_f$. Note that the decomposition $S_j = \bigcup_{l=-S_b}^{S_f} S_{jl}$ is induced by the function $f_{in}^d(m)$ or equivalently, by $f_{in}(m)$. Obviously, the codeword weight of the weight-2 information sequence $\mathbf{u}_{\pi_{ibp}^{-1}(m)\pi_{ibp}^{-1}(n)}$ is large, if $\mathbf{u}'_{mn}$ with $m \nsim n$. As we are concerned with $w_{2,min}$, only those weight-2 sequences with nonzero coordinate pairs in the set, $\{(m, n)|m \sim n \sim j, \text{ for some } j \text{ and } m, n \in S_{jl} \text{ for some } l\}$ have to be considered.

Assume that $\forall j, l$ all pairs $\{(m, n) \in S_{jl}\}$ satisfy the inequality $|m - n| > T_c(S_f + S_b + 1)$. For any pair $(m, n) \in S_{jl}, m < n$ and the associated interior set $V = \{m + 1, m + 2, \cdots, n - 1\}$, we have $|V| \geqslant T_c \cdot (S_f + S_b + 1)$. If $S_{jl} \cap V \neq \phi$, there exists a pair $(m', n') \in S_{jl}$, where $|m' - n'| < |m - n|$. Otherwise, if $S_{jl} \cap V = \phi$, by the pigeonhole principle [29], there exists a set $S_{pq}$ such that $|S_{pq} \bigcap V| \geqslant 2$, which implies that there is a pair $(m', n') \in S_{pq}$, where $|m' - n'| < |m - n|$.

As both cases lead to contradictions, we conclude that there exists a pair $(m, n) \in S_{jl}$ for some $j, l$, such that $|m - n| \leq T_c(S_f + S_b + 1)$. Since it is always possible to find $\pi_{intra}$ such that $|\pi_{ibp}^{-1}(m) - \pi_{ibp}^{-1}(n)| = T_c$, (17) and (18) then imply that $w_{2,min} \leq 2 + \alpha((T_c + T_c(S_f + S_b + 1))/T_c) + 2\beta$. When $S_f = S_b = S$, we have $w_{2,min} \leq 2 + 2\alpha(S + 1) + 2\beta$. ∎

As all data sequences are of finite length in practice, there are either no or not enough blocks for the first $S_b - 1$ and the last $S_f - 1$ blocks to perform either the complete backward or forward inter-block permutations. Therefore, we have to modify the IBP range for those blocks by reducing either the forward or the backward span. Assuming that there are $N \gg \max(S_f, S_b)$ blocks and denoting by $S_f(i)$ and $S_b(i)$ the forward and backward spans of the $i$th block, we require that for $0 \leq i < N$,

$$S_f(i) = \min\left(S_f, N - i + 1\right), \quad S_b(i) = \min\left(S_b, i\right). \tag{20}$$

*Theorem 2* is modified accordingly.

*Corollary 2:* For finite-length data sequences and a given inter-block interleaver, $\pi_{inter}$, whose spans are specified by (20), $\exists \; \pi_{intra}$ such that the corresponding IBPTC satisfies $w_{2,min} \leq 2 + \alpha \cdot \min(S_f + 2, S_b + 2) + 2\beta$, if $L > T_c \cdot \min(S_f + 1, S_b + 1)$.

*Theorem 2* and its *Corollary* indicate that lack of control on the intra-block interleaver imposes an upper-bound for $w_{2,min}$ an IBPTC can achieve. The coordinates of nonzero elements of the interleaved sequence $\mathbf{u}'_{ij}$ with $i \cong j$ will either remain in the same block or be in the different blocks with probabilities close to $1/(2S + 1)$ and $2S/(2S + 1)$ when considering all possible intra-block interleavers and assuming $S_f = S_b = S$. The resulting codewords for the latter case are very likely to have large weights while those for the former case have smaller weights with the worst-case weight of $2 + 2\alpha + 2\beta$ only.

To avoid generating low weight codewords for $\mathbf{u}_{ij}$, we first notice that (17) implies $\tilde{w}_{2,min} \geq 2 + \alpha(\delta_{min}/T_c) + 2\beta$. The IBP along with the intra-block interleaver determine the relation between $|i - j|$ and $|\pi_{ibp}(i) - \pi_{ibp}(j)|$, and their structures can be optimized to maximize $\delta_{min}$. For a pair of coordinates $(i, j) \in s_m$, if the integer-valued function $f_{in}^d(k)$ is injective and satisfies the locally-periodic property, $f_{in}^d(k) = f_{in}^d(k + nT_s)$, for $L|k|_L \leq n + kT_s < L(|k|_L + 1)$, where $T_s = S_f + S_b + 1$, then the requirements, $i \cong j$ and $\pi_{ibp}(i) \cong \pi_{ibp}(j)$ imply $||\pi_{ibp}(i) - \pi_{ibp}(j)||_{T_s} = 0$ and therefore $|\pi_{ibp}(i) - \pi_{ibp}(j)| \geq lcm(T_c, T_s)$, where $lcm(a, b)$ represents the least common multiple of $a$ and $b$. In other words,

*Lemma 2:* An IBPTC that uses a *Type III* inter-block interleaver satisfies

$$\min_{i \cong j, \pi_{ibp}(i) \cong \pi_{ibp}(j), (i,j) \in s_m} w_t(\mathbf{X}_{ij}) \geq 2 + \alpha\left\{\left[T_c + lcm(T_c, T_s)\right]/T_c\right\} + 2\beta. \tag{21}$$

If $T_c$ and $T_s$ are relative prime, then

$$\min_{i \cong j, \pi_{ibp}(i) \cong \pi_{ibp}(j), (i,j) \in s_m} w_t(\mathbf{X}_{ij}) \geq 2 + \alpha(S_f + S_b + 2) + 2\beta. \tag{22}$$

## C. Constraints on the intra-block interleavers

*Theorem 2* reminds us of the importance of a judicious choice of the intra-block interleaver. For the question of how to choose an intra-block interleaver whose associated $w_{2,min}$ is guaranteed to surpass the worst-case upper-bound of *Theorem 2*, *Lemma 2* gives only an unrefined answer. We need more elaborate constraints on the selection of the intra-block interleaver to avoid producing a $w_{2,min}$ smaller than that bound. In general, any one of the four conditions, $i \not\cong j$, $\pi_{ibp}(i) \not\cong \pi_{ibp}(j)$, $i \nsim j$, $\pi_{ibp}(i) \nsim \pi_{ibp}(j)$, is very likely to result in large $w_t(\mathbf{X}_{ij})$. However, there is still a small possibility that low weight codewords will be generated. Before presenting the requirements for eliminating these low weight codewords by using a proper intra-block permutation, we need to define a few new functions to facilitate our discussion.

We denote by $\mathbf{u}_k$ the weight-1 sequence whose only nonzero element is at coordinate $k$ and by $\widetilde{scrb}(\cdot)$ the RSC encoder that encodes a length-$L$ sequence and terminates at the all-zero state using proper tail-bits. Based on the above definitions, we further define, for $0 \leq i, j < L$

$$f_1(i,j) = \begin{cases} \alpha|i-j| + \beta, & \text{if } i \sim j \\ w_t(\widetilde{scrb}(\mathbf{u}_{ij})), & \text{otherwise} \end{cases} \tag{23}$$

$$f_2(i,j) = w_t(\widetilde{scrb}(\mathbf{u}_i)) + w_t(\widetilde{scrb}(\mathbf{u}_j)) \tag{24}$$

$$f_3(i,j) = \begin{cases} |i-j|, & \text{if } i \sim j \\ \infty, & \text{otherwise} \end{cases} \tag{25}$$

$$f_4(i,j) = \min(f_3(i,j), f_3(i,j+L), f_3(i,j-L)). \tag{26}$$

As the way a low weight codeword is generated depends on how the encoder terminates its state at the end of a block, we begin with the TP-IBPTCs.

### C.1 TP-IBPTC

For the class of TP-IBPTCs, a weight-2 input sequence $\mathbf{u}_{ij}, (i,j) \notin s_m$, can not generate an infinite-weight codeword because the encoder state is forced to be terminated at the all-zero state at the end of each block. On the other hand, low-weight codewords may be generated if

$$d_L(i) + d_L(j) + d_L(\pi_{ibp}(i)) + d_L(\pi_{ibp}(j)) < lcm(T_s, T_c) + T_c \tag{27}$$

where $d_L(n) = L - ||n||_L$ and in addition, (i) both $i, j$ and $\pi_{ibp}(i), \pi_{ibp}(j)$ are near the ends of different blocks, (ii) $i \cong j$, $\pi_{ibp}(i) \cong \pi_{ibp}(j)$ and both pairs lie close to the end of a block, or (iii) $i \ncong j$ or $\pi_{ibp}(i) \ncong \pi_{ibp}(j)$ but both pairs lie close to the end of a block. To avoid generating low weight codewords out of case (i), we require that

$$f_2(||i||_L, ||j||_L) + f_2\left(f_{ib}\left(L|i|_L + \pi_{block}(||i||_L)\right), f_{ib}\left(L|j|_L + \pi_{block}(||j||_L)\right)\right) \geq B(T_c, T_s) \tag{28}$$

where $B(T_c, T_s) = 2 + \alpha \left[T_c + lcm(T_c, T_s)/T_c\right] + 2\beta$. Similarly, for cases (ii)-(iii), $\pi_{block}$ must satisfy

$$f_1(||i||_L, ||j||_L) + f_1\left(f_{ib}\left(L|i|_L + \pi_{block}(||i||_L)\right), f_{ib}\left(L|j|_L + \pi_{block}(||j||_L)\right)\right) \geq B(T_c, T_s), \tag{29}$$

if $i \cong j$ and $\pi_{ibp}(i) \cong \pi_{ibp}(j)$, and

$$f_1(||i||_L, ||j||_L) + f_2\left(f_{ib}\left(L|i|_L + \pi_{block}(||i||_L)\right), f_{ib}\left(L|j|_L + \pi_{block}(||j||_L)\right)\right) \geq B(T_c, T_s) \tag{30}$$

if $i \cong j$ but $\pi_{ibp}(i) \ncong \pi_{ibp}(j)$, and

$$f_2(||i||_L, ||j||_L) + f_1\left(f_{ib}\left(L|i|_L + \pi_{block}(||i||_L)\right), f_{ib}\left(L|j|_L + \pi_{block}(||j||_L)\right)\right) \geq B(T_c, T_s) \tag{31}$$

if $i \ncong j$ but $\pi_{ibp}(i) \cong \pi_{ibp}(j)$.

The above conditions (28)-(31) are not too easy to meet but can be relaxed if we impose more constraints on $\pi_{inter}$. It is straightforward to show

*Lemma 3:* For an TP-IBPTC whose inter-block interleaver is of *Type IV*

$$\min_{i,j} w_t(\mathbf{X}_{ij}) \geq B(T_c, T_s)$$

if each element in the set $\Gamma_{T_s} = \{(i, j) : 0 \leq i, j \leq L - 1, ||\pi_{block}(i) - \pi_{block}(j)||_{T_s} = 0\}$ satisfies

$$\begin{aligned} f_2(i, j) + f_2(\pi_{block}(i), \pi_{block}(j)) &\geq B(T_c, T_s) \\ f_1(i, j) + f_1(\pi_{block}(i), \pi_{block}(j)) &\geq B(T_c, T_s) \end{aligned} \tag{32}$$

and $\forall\ (i, j) \notin \Gamma_{T_s}$ the following two inequalities are satisfied

$$\begin{aligned} f_1(i, j) + f_2(\pi_{block}(i), \pi_{block}(j)) &\geq B(T_c, T_s) \\ f_2(i, j) + f_1(\pi_{block}(i), \pi_{block}(j)) &\geq B(T_c, T_s). \end{aligned} \tag{33}$$

Starting with an arbitrary intra-block interleaver, say an *s*-random interleaver [10], we can apply the above criterion iteratively to find the smallest $L$ for a given component code such that $w_t(\mathbf{X}_{ij}) \geq B(T_c, T_s)$. When $L$ is large enough, e.g., $L > 2(T_c + lcm(T_c, T_s))$, the constraints imposed by the above lemma are relatively easy to meet, i.e., a $\pi_{intra}$ that satisfies these constraints is easy to find. For example, it just has to permute the bits near both ends of a block to those places far away from the ends.

## C.2 TB-IBPTC

Denote by $scrb_{tb}^l$ the tail-biting scrambler of length $l$, and define

$$
\begin{aligned}
S_1(l) &= \left\{ M = 2 + 2 scrb_{tb}^l(\mathbf{u}_i) \,|\, 0 \le i < l \right\} \\
S_2(l) &= \left\{ M = 2 + scrb_{tb}^l(\mathbf{u}_{ij}) \,|\, i \nsim j, i \nsim j \pm l, 0 \le i, j < l \right\} \quad (34) \\
S_k &= \bigcup_{l=k}^{\infty} \left[ S_1(l) \cup S_2(l) \right] \quad (35)
\end{aligned}
$$

and let $m_k$ be the smallest integer of the set $S_k$. Obviously, $\{m_k\}$ is a nondecreasing function of $k$. Denote the least integer $k$ such that $m_k \ge B(T_c, T_s)$ by $k_{min}$.

We observed that, for a TB-IBPTC whose block size $L \ge k_{min}$, a weight-2 sequence $\mathbf{u}_{ij}$ generates a codeword whose weight is less than the bound $B(T_c, T_s)$ only if $(i,j) \in \Gamma_{T_s}$ and $(i,j)$ satisfies the following conditions:

$$
\min \left\{ ||(|i-j|)||_{T_c}, ||(L-|i-j|)||_{T_c} \right\} = 0 \quad (36)
$$

$$
\min \left\{ ||(|\pi_{ibp}(i) - \pi_{ibp}(j)|)||_{T_c}, ||(L - |\pi_{ibp}(i) - \pi_{ibp}(j)|)||_{T_c} \right\} = 0. \quad (37)
$$

$$
\min \left\{ |i-j|, L - |i-j| \right\} + \min \left\{ |\pi_{ibp}(i) - \pi_{ibp}(j)|, L - |\pi_{ibp}(i) - \pi_{ibp}(j)| \right\} < lcm(T_c, T_s) + T_c. \quad (38)
$$

Such $(i,j)$ pairs will not exist if $\pi_{inter}$ is of *Type III* and the corresponding $\pi_{block}$ satisfies

$$
f_4(i,j) + f_4(f_{ib}(nL + \pi_{block}(i)), f_{ib}(nL + \pi_{block}(j))) \ge T_c + lcm(T_c, T_s), \ 0 \le i, j < L,
$$

$\forall \, n$ and $(i,j) \in \Gamma_{T_s}$. In manner similar to the TP-IBPTC case, the above constraint on $\pi_{block}$ can be further lessened when a *Type IV* inter-block interleaver is used. In summary,

*Lemma 4:* For a TB-IBPTC that uses a *Type IV* inter-block interleaver with a block length $L \ge k_{min}$, $w_{2,min} \ge B(T_c, T_s)$ if the corresponding $\pi_{block}$ satisfies

$$
f_4(i,j) + f_4(\pi_{block}(i), \pi_{block}(j)) \ge T_c + lcm(T_c, T_s). \quad (39)
$$

for all $(i,j) \in \Gamma_{T_s}$.

Note that in designing the interleaver for the classic turbo codes that use the identical tail-biting convolutional code as the component codes, one must also consider the constraint similar to *Lemma 4*.

## C.3 C-IBPTC

For the class of C-IBPTCs, we only have to consider $(i,j) \in s_m$. Low weight codewords are associated with those $(i,j)$ pairs whose combined pre-interleaved and post-interleaved distance, $|i-j| + |\pi_{ibp}(i) -$

$\pi_{ibp}(j)|$, is small. The upper-bound promised by *Theorem 2* can be achieved if

$$f_4(i, j) + f_4(f_{ib}(nL + \pi_{block}(i)), f_{ib}(nL + \pi_{block}(j))) \geq T_c + lcm(T_c, T_s), \tag{40}$$

for all $n$ and $(i, j) \notin \Gamma_{T_s}$, if $\pi_{inter}$ is of *Type III*.

The constraint (40) is used to ensure that the pair $(\pi_{ibp}(i), \pi_{ibp}(j))$ though in different blocks (since $(i, j) \notin \Gamma_{T_s}$) are separated by a large distance.

In analogy to the case of the TB-IBPTC, the constraint on $\pi_{block}$ can be relaxed if the corresponding $\pi_{inter}$ is more restricted. It is easy to show

*Lemma 5:* For a C-IBPTC that uses a *Type IV* inter-block interleaver, if the associated $\pi_{intra}$ is such that for all $(i, j) \notin \Gamma_{T_s}$,

$$f_4(i, j) + f_4(\pi_{block}(i), \pi_{block}(j)) \geq T_c + lcm(T_c, T_s),$$

then $w_{2,min} \geq B(T_c, T_s)$.

### C.4 Finite-length IBPTCs

To accommodate the IBP ranges defined by (20) for the finite-length inputs, the range of $f_{in}(i)$ and the period $T_s$ of a *Type II* or *III* inter-block interleaver must be adjusted according to

$$\max\left(-S_b, -|i|_L\right) \leq f_{in}(i) \leq \min\left(S_f, N - 1 - |i|_L\right), \tag{41}$$

$$T_s(n) = \begin{cases} n + S_f + 1, & \text{if } 0 \leq n < S_b \\ N - n + S_b, & \text{if } N - S_f \leq n < N \\ S_f + S_b + 1, & \text{otherwise} \end{cases}, \tag{42}$$

where $0 \leq n < N$. Even with the above modifications, low-weight codewords can still be generated for some weight-2 input sequences. A simple solution is to adjust the block lengths of the beginning $S_b$ and the last $S_f$ blocks such that the block length of the $i$th block satisfies

$$nT_s(n) \leq L(n) < nT_s(n) + T_s(n), \tag{43}$$

for some $n$.

*Lemma 6:* For an $N$-block IBPTC whose block lengths $L(n)$ are given by (43) and whose $\pi_{inter}$ is of *Type IV* IBP with local interleaving periods defined by (42),

$$\min_{i \cong j, \pi_{ibp}(i) \cong \pi_{ibp}(j)} w_t(\mathbf{X}_{ij}) \geq 2 + \alpha \left\lceil \frac{T_c + lcm(T_c, T_s(n))}{T_c} \right\rceil + 2\beta. \tag{44}$$

Finite-length versions of *Lemmas 3-4* can also be established if the block length and the corresponding IBP rule meet the requirements stated in the above lemma. For a C-IBPTC, however, $\pi_{block}$ needs to satisfy the additional requirement that for all $0 \leq i, j < L$ such that $||\pi_{block}(i) - \pi_{block}(j)||_{T_s} = 0$

$$f_1(i, j) + f_1(\pi_{block}(i), \pi_{block}(j)) \geq 2 + \alpha \left[ \frac{T_c + lcm(T_c, S_b + 1)}{T_c} \right] + 2\beta. \tag{45}$$

When this requirement is also met then we have

$$\min_{i,j} \; w_t(\mathbf{X}_{ij}) \geq 2 + \alpha \left[ \frac{T_c + lcm(T_c, S_b + 1)}{T_c} \right] + 2\beta. \tag{46}$$

The above discussion shows that the *Type IV* inter-block interleavers do possess some desired properties and should be used in conjunction with a proper intra-block interleaver. Hokfelt *et al.* [7] showed that, as the correlation function of the extrinsic output is exponentially decayed, the interleaver should separate neighboring bits as far as possible. The local periodicity requirement of the *Type IV* interleavers is consistent with this intuition and let bits or samples within the neighborhood of $S_f + S_b$ blocks be moved to the different blocks.

## V. Upper-bounds of codeword weights for weight-2 and weight-4 input sequences

This section derives upper-bounds for the weights of IBPTC codewords associated with weight-2 and weight-4 input sequences. These upper-bounds are valid for all intra- and inter-block interleavers.

Recall that *Lemma 1* implies that, the minimum codeword weight, $w_{2,min}$, for the weight-2 input sequences whose coordinates $(i, j)$ of nonzero elements satisfy $i \sim j$ and $\pi(i) \sim \pi(j)$ is upper-bounded by

$$w_{2,min} \leq 2 + \alpha \cdot \left( \frac{|i - j| + |\pi(i) - \pi(j)|}{T_c} \right) + 2\beta, \tag{47}$$

where it is understood that the constants $\alpha$ and $\beta$ might not have the same values as those of (17). A bound much tighter than (47) can be obtained by applying the approach suggested by Breiling [28] who partitions the coordinates set associated with both pre-interleaved and post-interleaved sequences into equivalence classes induced by the equivalent relation "$\sim$". Each equivalence class is further divided into subsets $F_z = \{z + mT_c, m = 0, 1, \cdots, |F_z| - 1\}$, where $z$ is the smallest index in $F_z$.

An output (parity) sequence will be of finite weight if the coordinate pair $(i, j)$ associated with the weight-2 input sequence $\mathbf{u}_{ij}$ belongs to the same equivalence class. The parity sequence weight is small if the pair $(i, j)$, besides being in the same equivalence class, are in the proximity of each other, i.e., if $(i, j) \in F_z$ for some $z$ and the width of $F_z = (|F_z| - 1)T_c$ is small.

To avoid generating low-weight codewords, therefore, an optimal interleaver should send any pair of coordinates in a given subset to different equivalent classes and, if that is not possible, to different subsets or at least to two far-apart coordinates within a subset. Let $F_z^{(m)}$ and $\Lambda_m$, $(0 \leq z < \Lambda_m)$ be the subsets and the number of subsets associated with the coordinates of the $m$th component encoder input sequence. The cardinalities of the $\Lambda_m$ subsets differ at most by 1, i.e., $|F_z^{(m)}| = \lfloor L/\Lambda_m \rfloor$ or $\lfloor L/\Lambda_m \rfloor + 1$. Invoking the aforementioned pigeonhole principle, Breiling showed that if the pair $(\Lambda_1, \Lambda_2)$ is such that $\lceil L/\Lambda_1 \rceil > \Lambda_2$ then any interleaver would map a pair of coordinates $(i, j)$ that lies in the same subset $F_z^{(1)}$ to $(\pi(i), \pi(j))$ which also belongs to an identical subset $F_{z'}^{(2)}$, resulting in

$$w_t(\mathbf{X}_{ij}) \leq 2 + \alpha \left( \left\lceil \frac{L}{\Lambda_1} \right\rceil + \left\lceil \frac{L}{\Lambda_2} \right\rceil - 2 \right) + 2\beta. \tag{48}$$

Minimizing the right hand side of the above inequality with respect to the the pair $(\Lambda_1, \Lambda_2)$, Breiling then obtained a very tight upper-bound.

### A. Upper-bound for weight-2 input sequences

It is clear that, given the same set of parameters $\{L, \Lambda_i, T_c, |F_z^{(i)}|\}$, an IBP interleaver has subsets within its span to choose from for placing members of the set $\{\pi(i), i \in F_z^{(1)}\}$, for some $0 \leq z < L$. Thus, assuming a large enough block size $(L)$, the priority of an optimal IBP rule in permuting coordinates of the same equivalence class follows the order: (i) to different blocks, (ii) to different equivalent classes of the same block, (iii) to different subsets of the same equivalence class, and finally, (iv) to far-apart coordinates within the same subset. Obviously, the partition of an equivalence class into subsets plays a pivotal role in optimizing an IBP rule. With a minimum loss of generality, we assume $||\Lambda_1||_M = ||\Lambda_2||_M = 0$, $M = T_s T_c$, where $T_s = 2S + 1$. Given these parameter values, we consider the following (subset) partition.

$$F_i^{(k)} = \begin{cases} \left\{ ||i||_M + [i - ||i||_M] \left\lceil \frac{L}{\Lambda_k} \right\rceil + Mj : \ 0 \leq j < \left\lceil \frac{L}{\Lambda_k} \right\rceil \right\}, \ \text{where } 0 \leq i < ||L||_{\Lambda_k} - ||L||_M. \\[2mm] \left\{ ||i||_M + [||L||_{\Lambda_k} - ||L||_M] \left\lceil \frac{L}{\Lambda_k} \right\rceil + [i - ||i||_M - ||L||_{\Lambda_k} + ||L||_M] \left\lfloor \frac{L}{\Lambda_k} \right\rfloor + Mj : \ 0 \leq j < \left\lfloor \frac{L}{\Lambda_k} \right\rfloor \right\}, \\[1mm] \quad \text{where } ||L||_{\Lambda_k} - ||L||_M \leq i < \Lambda_k - M. \\[2mm] \left\{ ||i||_M + [||L||_{\Lambda_k} - ||L||_M] \left\lceil \frac{L}{\Lambda_k} \right\rceil + [\Lambda_k - ||L||_{\Lambda_k} - M + ||L||_M] \left\lfloor \frac{L}{\Lambda_k} \right\rfloor + Mj : \ 0 \leq j < \left\lceil \frac{L}{\Lambda_k} \right\rceil \right\}, \\[1mm] \quad \text{where } \Lambda_k - M \leq i < \Lambda_k - (M - ||L||_M). \\[2mm] \left\{ ||i||_M + [||L||_{\Lambda_k} - ||L||_M] \left\lceil \frac{L}{\Lambda_k} \right\rceil + [\Lambda_k - ||L||_{\Lambda_k} - M + ||L||_M] \left\lfloor \frac{L}{\Lambda_k} \right\rfloor + Mj : \ 0 \leq j < \left\lfloor \frac{L}{\Lambda_k} \right\rfloor \right\}, \\[1mm] \quad \text{where } \Lambda_k - (M - ||L||_M) \leq i < \Lambda_k. \end{cases} \tag{49}$$

An exemplary partition of (49) is shown in Fig. 6 where the integers represent the coordinates of either an input or output sequence and each row consists of three segments with a segment representing a subset of size 3 or 2. The IBP rule sends bits in rows labelled by different capital letters to different blocks while those in the same row are interleaved to the same block.

By using an argument similar to that leading to (48) and invoking the partition of (49) along with the permutation rule (i)-(iv) mentioned at the beginning paragraph of this subsection, we obtain

*Theorem 3:* For the class of IBPTCs, the minimum codeword weight $w_{2,min}$ for weight-2 input sequences is upper-bounded by

$$w_{2,min} \leq 2 + \alpha \left( \min_{(\Lambda_1, \Lambda_2)} \left\{ \left\lceil \frac{T_s L}{\Lambda_1} \right\rceil + \left\lceil \frac{T_s L}{\Lambda_2} \right\rceil \right\} - 2 \right) + 2\beta \qquad (50)$$

where $(\Lambda_1, \Lambda_2) \in D \times D$, $D = \{1, 2, \cdots, \lceil L/M \rceil - 1\}$, if $L > MT_c$ and

$$\left\lceil \frac{L}{\Lambda_1} \right\rceil > \frac{\Lambda_2}{T_s}.$$

When $\Lambda_1 = \Lambda_2$, we have

$$w_{2,min} \leq 2 + 2\alpha \frac{T_s L}{\sqrt{T_s L} - T_s T_c} + 2\beta, \qquad (51)$$

.

*Proof:* (51) follows directly from the partition (49) and the optimal periodic IBP. The corresponding interleaver results in bound-achieving codewords $\mathbf{X}_{ij}$, $(i, j) \in s_m$, when $\lceil \frac{L}{\Lambda_1} \rceil > \frac{\Lambda_2}{T_s}$. Hence

$$\min_{(\Lambda_1, \Lambda_2)} \left\{ \left\lceil \frac{T_s L}{\Lambda_1} \right\rceil + \left\lceil \frac{T_s L}{\Lambda_2} \right\rceil \right\} = \min_{\Lambda_1} \left\{ \left\lceil \frac{T_s L}{\Lambda_1} \right\rceil + \min_{\lceil \frac{L}{\Lambda_1} \rceil > \frac{\Lambda_2}{T_s}} \left\lceil \frac{T_s L}{\Lambda_2} \right\rceil \right\}$$

$$\leq \min_{\Lambda_1} \left\{ \left\lceil \frac{T_s L}{\Lambda_1} \right\rceil + \left\lceil \frac{L}{T_c (\lceil \frac{L}{T_c \Lambda_1} \rceil - 1)} \right\rceil \right\} \qquad (52)$$

The upper-bound (50) can be rewritten as

$$w_{2,min} \leq 2 + \alpha \left( \min_{\Lambda_1} \left\{ \left\lceil \frac{T_s L}{\Lambda_1} \right\rceil + \left\lceil \frac{L}{T_c (\lceil \frac{L}{T_c \Lambda_1} \rceil - 1)} \right\rceil \right\} - 2 \right) + 2\beta \qquad (53)$$

If we choose $(\Lambda_1, \Lambda_2) = (\Lambda_0, \Lambda_0)$ with $\Lambda_0 = M(\lceil \frac{\sqrt{T_s L}}{M} \rceil - 1)$, i.e., $\Lambda_0$ is a multiple of $M$ and $\Lambda_0^2 < T_s L$, then (50) implies

$$w_{2,min} \leq 2 + 2\alpha \left( \left\lceil \frac{T_s L}{M \cdot (\lceil \frac{\sqrt{T_s L}}{M} \rceil - 1)} \right\rceil - 1 \right) + 2\beta$$

$$\leq 2 + 2\alpha \left( \left\lceil \frac{T_s L}{\sqrt{T_s L} - M} \right\rceil - 1 \right) + 2\beta$$

$$< 2 + 2\alpha \left( \frac{T_s L}{\sqrt{T_s L} - T_s T_c} \right) + 2\beta \qquad (54)$$

Theorem 3 implies that $w_{2,min}$ grows linearly with $\sqrt{T_s L}$ when $L$ is large.

## B. Upper-bound for weight-4 input sequences

Let the coordinates of nonzero elements of a weight-4 input sequence be $(i, j, k, l)$, where $i < j < k < l$. If we divide these coordinates and their permuted positions respectively into two pairs each according to their natural order, i.e., $(i, j), (k, l)$ and say, $(\pi(i), \pi(k)), (\pi(j), \pi(l))$, then a low-weight codeword results if each pair belongs to the same subset. More specifically, the minimum codeword weight, $w_{4,min}$, for weight-4 input sequences whose nonzero coordinates $(i, j, k, l)$ are such that $i \sim j, k \sim l, \pi(i) \sim \pi(k), \pi(j) \sim \pi(l)$ satisfies

$$w_{4,min} \leq 4 + \alpha \cdot \left( \frac{|i-j| + |k-l| + |\pi(i) - \pi(k)| + |\pi(j) - \pi(l)|}{T_c} \right) + 2\beta \tag{55}$$

or

$$w_{4,min} \leq 4 + \alpha \cdot \left( \frac{|i-j| + |k-l| + |\pi(i) - \pi(j)| + |\pi(k) - \pi(l)|}{T_c} \right) + 2\beta. \tag{56}$$

if $(i, j, k, l)$ are such that $i \sim j, k \sim l, \pi(i) \sim \pi(j), \pi(k) \sim \pi(l)$.

These upper-bounds are obtained by considering the three pre- and post-interleaving distributions of the 4-tuple $(i, j, k, l)$ shown in Fig. 7 (a)-(c). These three are the distributions that most likely lead to low-weight codewords. There are other candidate distributions (e.g., Fig. 7 (d)) but the corresponding upper bounds are likely to be larger that those given by (55) and (56).

Following an approach similar to that of [26] and taking into account the extra degrees of freedom offered by an IBP interleaver, we obtain

*Theorem 4:* The IBPTC minimum codeword weight for weight-4 input sequences is upper-bounded by

$$w_{4,min} \leq 4 + 2\alpha \left( \min_{(\Lambda_1, \Lambda_2)} \left\{ \left\lceil \frac{T_s L}{\Lambda_1} \right\rceil + \left\lceil \frac{T_s L}{\Lambda_2} \right\rceil \right\} - 2 \right) + 4\beta \tag{57}$$

when $(\Lambda_1, \Lambda_2) \in D \times D$, where $D = \{1, 2, \cdots, \lfloor L/2 \rfloor\}$, satisfies (i) $\Lambda_1 \binom{\Omega}{2} > \binom{\Lambda_2}{T_s^2}$, (ii) $\frac{(T_s - k)}{T_s} \cdot \Lambda_1 \Omega^2 > \left( \frac{\Lambda_2}{T_s} \right)^2$, and (iii) $||\Lambda_i||_M = 0$, $i = 1, 2$, where $\Omega = \lfloor \frac{L}{\Lambda_1} \rfloor$ and $k = 1, 2, \cdots, T_s - 1$. Moreover, for the special case, $\Lambda_1 = \Lambda_2$ and if $L > \frac{10}{3} T_s^3 + T_s^2 - \frac{T_s}{3}$ and $T_S > 1$ the upper-bound yields the compact expression

$$w_{4,min,ibp} \leq 4 + 4\alpha \frac{T_s L}{C - T_s T_c} + 4\beta, \tag{58}$$

where

$$C = \frac{T_s + 2T_s^2}{3} + \sqrt[3]{-\frac{q_1}{2} + \sqrt{\left(\frac{p_1}{3}\right)^3 + \left(\frac{q_1}{2}\right)^2}} + \sqrt[3]{-\frac{q_1}{2} - \sqrt{\left(\frac{p_1}{3}\right)^3 + \left(\frac{q_1}{2}\right)^2}} \tag{59}$$

$$p_1 = 3T_s^2 L - \frac{1}{3}(T_s + 2T_s^2)^2, \tag{60}$$

$$q_1 = -T_s^2 L^2 + (T_s^3 + 2T_s^4)L - \frac{2}{27}(T_s + 2T_s^2)^3. \tag{61}$$

*Proof:* See Appendix B. ∎

Again, we observe that for large $L$, the upper-bound grows linearly with $(T_s L)^{\frac{1}{3}}$. The minimum codeword weights associated with weight-2 and weight-4 input sequences are upper-bounded by the increasing functions of $T_s L$.

### C. Interleaving gain comparison

As shown in Section III, except for the first output block, an IBPTC decoder yields an inter-block decoding delay the same as that of a classic TC with the same block size. However, the encoding delay or the SRID of an IBPTC is $(1 + S)$ times larger. Let $w_{2,min,block}$ and $w_{4,min,block}$ be the minimum codeword weights associated with weight-2 and weight-4 input sequences of classic TCs with block size $(S + 1)L$, then we have [28]

$$w_{2,min,block} \leq 2 + 2\alpha \frac{(S+1)L}{\sqrt{(S+1)L - T_c}} + 2\beta \tag{62}$$

$$w_{4,min,block} \leq 4 + 4\alpha \frac{(S+1)L}{((S+1)L-1)^{\frac{2}{3}} - ((S+1)L-1)^{\frac{1}{3}} + 1 - T_c} + 4\beta. \tag{63}$$

Comparing the above equations with (51) and (58) and noting that $T_s = 2S + 1$, we conclude that, as far as weight-2 and weight-4 input sequences are concerned, a 'good' IBPTC can bring about improvement factors of $\left(2 - \frac{1}{S+1}\right)^{\frac{1}{2}}$ and $\left(2 - \frac{1}{S+1}\right)^{\frac{1}{3}}$, respectively.

## VI. IBPTC ARCHITECTURE AND ALGORITHMS

### A. Implementation concern of IBPI

We have defined a swap interleaver as one such that $\forall i \ \pi(i) = \pi^{-1}(i)$. As an IBPI moves bits or symbols in a given block to positions within itself and those in the neighboring $S_f + S_b$ blocks, the associated interleaver normally requires at least $(S_f + S_b + 1)L$ units of memory (see Fig. 8(a)), where the number of bits per unit depends on the system's precision requirement. However, if we use a symmetric $(S_f = S_b = S)$ swap interleaver, then only $(S + 1)L$ units of memory are needed for temporary interleaving or deinterleaving storage. As shown in Fig. 8(b), we do not have to move those forwardly-permuted symbols until after all earlier (backward) blocks have been filled by interleaved (or deinterleaved) symbols and after their contents have been dumped. Moreover, a symmetric swap interleaver has the same interleaving and deinterleavering structure and can be implemented by single permutation table or algorithm. These advantages of symmetric swap IBPIs will still be maintained when we consider the implementation of the combined intra- and inter-block permutations. An IBPI using the swap structure has only to perform memory content swapping between current block and the backward blocks. Furthermore, if $\pi_{inter}$ is a *Type IV* interleaver, the only IBP operation is simply $m \rightarrow m - nL$, where $n \in \{1, 2, \cdots, S\}$.

Theorems 1 and 2 give us some guidelines for designing an IBP algorithm. In the previous paragraph, we show that the IBP with the swap structure has an implementation edge. Shown in Table I is a symmetric IBPI with $S = S_f = S_b$ and SRID $= (S + 1)L$. It can be easily seen that

*Corollary 3:* The algorithm in Table I satisfies the requirements of both *Type IV* and *Type V* interleavers.

### B. Modified semi-random interleaver

Semi-random interleavers [10] are designed to eliminate "short cycles" that send two close-by bits to the vicinity of each other after interleaving. These interleavers are, however, originally designed to work in the block interleaving setting, therefore they can not avoid two new classes of short cycles arising in TB-IBPTCs and C-IBPTCs. A tail-biting convolutional code begins and ends at the same state, hence if two close-by bits in a block are respectively intra-block permuted to the beginning and the ending parts of that block, and if the two bits remain in the same block after the IBP interleaving, a short cycle will result as the proposed IBP does not alter their relative positions within a block. For the class of C-IBPTCs, we also want to prevent similar intra-block interleaving results because the IBPI may send such a pair to the ending and beginning parts of two neighboring blocks. We therefore modify the constraint of [10] as

$$d_{min}(i, j) + d_{min}(\pi(i), \pi(j)) > S_2, \ 0 \leq i, j < L \tag{64}$$

where $d_{min}(i, j) = \min(|i-j|, L-|i-j|)$. This new constraint excludes the possibility that two symbols at the beginning and the ending parts of a block would remain there after the interleaving.

## VII. Simulation Results

### A. Error probability performance

Computer simulation results reported in this section use the RSC code of the 3GPP standard, $G(D) = \frac{1+D+D^3}{1+D^2+D^3}$ [27], the interleaver of the same standard or the modified semi-random interleaver of (64) for the intra-block permutation while the IBP follows the algorithm of Table I. Each simulation run consists of 1000 blocks. We use the Log-MAP or MAX-Log-MAP algorithms for decoding classic TCs and TP-IBPTCs, the sliding-window Log-MAP or the sliding-window MAX-Log-MAP algorithms for decoding TB-IBPTCs and C-IBPTCs. In most cases, we compare the performance of classic TCs and IBPTCs under the assumption that either both codes have the same encoding delay, SRID or they have the same average IBDD. As discussed in Section II, the latter implies that both codes use a single APP decoder and identical block size $L$, and the former case implies that the classic TC uses a block size of $(1 + S)L$ while the IBPTC has a block size of $L$.

Figs. 9 and 10 show the BER performance of rate 1/3 turbo coded systems with 10 iterations and Log-MAP algorithm. The interleaver parameter values for the IBPTC are $L = 402$, $S = 1$ or $L = 265$, $S = 2$. Compared with the performance of the classic TC with $L = 400$, the IBPTCs yield 0.7–0.9 dB performance gain at BER=$10^{-4}$ and 1.0–1.2 dB gain at BER=$10^{-6}$. When both codes have the same SRID, the IBPTC provides 0.4–0.6 dB performance gain at BER between $10^{-4}$ and $10^{-6}$.

Figs 11 and 12 show the BER performance of rate 1/2 turbo coded systems. The MAX-Log-MAP algorithm is used in this example. We compare the performance of the classic TC with $L = 1320$ and the IBPTCs with $L = 660$, $S = 1$ and $L = 440$, $S = 2$. Using $L = 660$, $S = 1$ and the 3GPP interleaver as the intra-block interleaver, the IBPTCs have 0.4–0.45 dB and 0.3 dB gain at BER=$10^{-5}$ and $10^{-6}$, respectively. For other cases, the IBPTCs give 0.4–0.45 dB gain at BER=$10^{-5}$ and 0.4-0.6 dB gain (except for the case TP-IBPTC with $L = 440, S = 2$) at BER=$10^{-6}$. It is clear that the IBPTCs outperform the conventional TCs with nearly the same SRID. Furthermore, the proposed modified s-random interleaver outperforms the 3GPP defined interleaver, especially when the interleaver span is small ($S = 1$).

These figures reveal that the proposed IBPTCs yield superior performance, sharper slope of the BER curve at the waterfall region and lower error floor when compared with the corresponding performance curves of the classic TCs for a variety of different code rates and decoding algorithms. The improvement is more impressive for smaller SRIDs, and with the same SRID, a larger interleaver span ($S$) leads to better performance.

Fig. 13 shows the BER performance of rate 1/3 IBPTCs that use the 3GPP defined interleaver as the intra-block interleaver. Either the Log-MAP algorithm or the Log-MAP algorithm is used and 15 decoding iterations is assumed. All these IBP parameter values, $(L, S) = (660, 1), (440, 2)$ or $(330, 3)$, give the same SRID of 1320 samples. The performance is consistent with our prediction: the larger the interleaver span is, the better the system performance becomes. The performance deteriorates when the period of encoder, $T_c$, and the period of the IBPI, $T_s$, are the same. For this case the lower-bound of (21) becomes $2(1 + \alpha + \beta)$ which is much smaller than the corresponding upper-bound given in *Theorem 2*. By contrast, the two bounds are much closer if $T_c \neq T_s$ and both bounds give identical value if $T_c$ and $T_s$ are relative prime.

Finally, we want to show that the IBPTC requires an interleaver latency much smaller than that of classic TCs with similar BER performance. Fig. 14 shows the BER performance of rate 1/3 turbo coded systems that employ 10 decoding iterations and the Log-MAP algorithm. All the interleavers are taken from the 3GPP defined interleaver. The average interleaver and deinterleaver latency of the IBPTCs is about 800. It is observed that the performance of the IBPTCs is bounded by those of turbo codes with block size $L = 2800$ and $L = 3600$. In other words, an IBPTCs achieves BER performance similar to that of a classic TC which requires an interleaving latency 3.5 to 4.5 times longer.

All these figures show that the TB-IBPTC has the best performance, followed by the C-IBPTC and then the TP-IBPTC.

*B. Covariance and convergence behavior*

Fig. 15 shows the covariance behavior for both IBPTC and classic TC with the same SRID, where the IBPI has $S = 1$ and SRID $= 800$ and the interleaving depth for the classic TC is $L = 800$. It indicates that the covariance is small for the IBPTC even at SNR $= 0.5$ dB while much higher covariance is observed for the classic TC at much higher SNR. The IBP collects extrinsic information from farther and farther away as the number of iterations increases and we have expected that it results in smaller covariance.

Two similar techniques have been proposed to study the convergence behavior of iterative decoding schemes, namely, the extrinsic information transfer chart (EXIT chart) of [23], [22] and the extrinsic information SNR evolution chart of [21]. The latter is a simplified version of Richardson's density evolution approach [24]. As the extrinsic information in the iterative decoding can be approximated by a Gaussian random variable, the evolution of the corresponding probability density as a function of iteration can be characterized by the SNR evolution where SNR is defined as the squared mean to variance ratio of the density.

Fig. 17 compares the EXIT behavior of our proposal and the classic TC with the same SRID. The IBPTC yields mutual information almost equal to one at SNR = 0.5 and 1.0 dB, but the classic TC needs SNR = 2.0 dB to achieve the same performance. The SNR evolution chart shown in Fig. 16 exhibits similar behavior of the two codes, all indicating the proposed IBPTC gives superior performance. Both figures also reveal that our code has a much faster convergence speed. The much larger step of the IBPTC curves means the associated APP decoder generates more information or extrinsic information with larger signal to noise ratio for the next stage decoder. Such a trend has been expected when we examine the factor graph structure of the IBPTC in Fig. 2.

## VIII. Conclusion

We present a class of IBP interleavers that enables an iterative decoder to collect information from a large span of neighboring samples with a bounded SRID or average IBDD. We derive the worst-case codeword weight upper bound for the weight-2 input sequences and provide constraints on the selection of the associated intra-block interleaver. The codeword weight upper-bounds for the weight-2 and weight-4 input sequences, when we have the freedom to select both the inter- and intra-block interleavers, are also given. It is shown that these bounds are better than those of the classic TCs with the same SRID. Our analysis also indicates that an IBP rule that possesses some regularities like periodicity and symmetry is likely to be a good IBP though global optimality has not been established.

Using some of the properties and bounds we derived as design guidelines, we propose a simple IBP algorithm, suggest a modified semi-random intra-block interleaver and address some implementation and hardware architecture issues. Simulation results based on the 3GPP standard turbo component code show that the IBPTCs provide $0.3 \sim 1.2$ dB performance gain. The performance curves have sharper slopes in the waterfall region with respect to those of the classic TCs with the same SRID or average IBDD. The class of IBPTCs achieve the same performance as that of the classic TC with a much a reduced SRID.

The class of proposed IBPTCs provides flexibility and tradeoffs that are not found in the classic TCs. In particular, it possesses some desired features that suit high data rate applications naturally. Since it can use any existing block-wise interleaver as its intra-block interleaver, the encoder/decoder structure is backward compatible in the sense that the special case $S = 1$ degenerates to the classic TC structure.

# Appendix

## Proof of Theorem 4

We first notice that, besides those finite weight codewords resulting from termination, as illustrated in Fig. 7, there are three conditions under which a weight-4 input sequence of an IBPTC will generate a finite-weight codeword. In Case (a), the codeword consists of two finite-weight segments (in different blocks) generated respectively by two weight-2 input sequences and thus the corresponding codeword weight upper-bound is simply twice that given in *Theorem 3*. Case (b) considers the situation when two pairs of coordinates from $F_i^{(1)}$ and $F_j^{(1)}$ of either the same block or different blocks are permuted to the same block with one coordinate from each pair mapped to two subsets $F_k^{(2)}$ and $F_l^{(2)}$, where the pair $(k, l), k \neq l$ belongs to the same equivalence class while the remaining two coordinates mapped to another two subsets $F_m^{(2)}$ and $F_n^{(2)}$ with $m \neq n$ in another equivalence class. Case (c) is similar to Case (b) except that the two subsets that contain the two permuted pairs are in different blocks.

Note that if $k = l$ and $m = n$ then the both cases will result in a codeword weight upper-bound similar to that obtained in [28]. But this is impossible as coordinates from different blocks will not be mapped into coordinates in the same subset (defined by (49)) by an optimal interleaver. This is because the spatial symmetric structure of a classic TC implies that, for every input sequence $\mathbf{u}$ of the code $\mathbf{X}$ that uses the interleaver $\pi$, $\exists \mathbf{u}'$ such that the codewords generated by $(\mathbf{u}, \pi)$ and $(\mathbf{u}', \pi^{-1})$ have identical weight. This observation and the fact that both component encoder outputs, $\mathbf{x}^1$ and $\mathbf{x}^2$, contribute equally to the resulting codeword weight suggest that $\pi$ and $\pi^{-1}$ have the same effect on the weight distribution, and that optimizing the deinterleaver rule results in the same mapping as the optimal interleaver.

Since we have to consider the scenario $k \neq l$ and $m \neq n$ only, the worst case occurs when both $|k - l|$ and $|m - n|$ are less than $T_s T_c$. In other words, Cases (b) and (c) concern the situation in which the pairs $(\pi_{ibp}(x), \pi_{ibp}(w))$ and $(\pi_{ibp}(y), \pi_{ibp}(z))$ belong to distinct supersubsets where a supersubset $\tilde{F}_j^{(2)}$ consists of $M/T_s = T_c$ consecutive subsets of the same equivalence class. Each block therefore has $\frac{\Lambda}{T_s}$ supersubsets, and $F_k^{(2)}$ and $F_l^{(2)}$ are in the same supersubset $\tilde{F}_j^{(2)}$ if $|k|_M = |l|_M = j$ and $||k - l||_{T_s} = 0$, or equivalently, $\tilde{F}_p^{(i)} = \bigcup_{k=0}^{T_s-1} F_{||j||_{T_c} + kT_c + |j|_{T_c} M}^{(i)}$, $i = 1, 2$.

Let $\Lambda_1$ and $\Lambda_2$ be the number of coordinates subsets per block for the input and permuted sequences. The subset partition rule, (49), implies that $\Omega \leq |F_i^{(j)}| \leq \Omega + 1$, where $\Omega = \lfloor \frac{L}{\Lambda_i} \rfloor$. For Case (b), each subset has either $\binom{\Omega+1}{2}$ or $\binom{\Omega}{2}$ distinct coordinates pairs and each block has at least $\Lambda_i \binom{\Omega}{2}$ such pairs. Our IBP interleaver maps $\frac{\Lambda_1}{T_s}$ sets of coordinates to each block within its span, or equivalently, a block "receives" coordinates from $T_s$ neighboring blocks. The optimal IBP rule would map a pair of coordinates in the same subset to different equivalent classes or blocks and, when this is not possible, to different supersubsets of the same block.

A pair of coordinates $(i, j)$ in $F_i^{(1)}$ can be mapped to any one of the $\binom{\frac{\Lambda_2}{T_s}}{2}$ pairs of distinct supersubsets $\tilde{F}_j^{(2)}, \tilde{F}_k^{(2)}, j \neq k$ of a neighboring block. A periodic IBP requires that at least $T_s \frac{\Lambda_1}{T_s} \binom{\Omega}{2}$ distinct pairs of coordinates from $T_s$ neighboring blocks be permuted to the same block. The pigeonholes principle implies that Case (b) will occur if

$$\Lambda_1 \binom{\Omega}{2} > \binom{\frac{\Lambda_2}{T_s}}{2}. \tag{A.1}$$

For Case (c) the pairs $(\pi_{ibp}(x), \pi_{ibp}(w))$ and $(\pi_{ibp}(y), \pi_{ip}(z))$ are in two distinct blocks. If the two distinct blocks are separated by $k$ blocks ($k = 1$ means they are two successive blocks), then $(\pi_{ibp}(x), \pi_{ibp}(w)), (\pi_{ibp}(y), \pi_{ip}(z))$ are mapped from $T_s - k$ neighboring blocks in which each block contains $\frac{\Lambda_1}{T_s}$ supersubsets and each supersubset has at most $(\Omega + 1)^2$ and at least $\Omega^2$ coordinates pairs to the two designated blocks. Therefore, finite weight codewords result if

$$(T_s - k)\frac{\Lambda_1}{T_s}\Omega^2 > \left(\frac{\Lambda_2}{T_s}\right)^2 \tag{A.2}$$

and we obtain upper-bounded

$$w_{4,min,ibp} \leq 4 + 2\alpha \cdot \left( \min_{(\Lambda_1,\Lambda_2)} \left\{ \left\lceil \frac{T_s \cdot L}{\Lambda_1} \right\rceil + \left\lceil \frac{T_s \cdot L}{\Lambda_2} \right\rceil \right\} - 2 \right) + 4\beta \tag{A.3}$$

where $(\Lambda_1, \Lambda_2)$ are subject to the constraints, (C1): $||\Lambda_1||_M = ||\Lambda_2||_M = 0$, (C2): $\Lambda_1 \binom{\Omega}{2} > \binom{\frac{\Lambda_2}{T_s}}{2}$, and (C3): $\frac{(T_s-k)}{T_s}\Lambda_1\Omega^2 > \left(\frac{\Lambda_2}{T_s}\right)^2$. Since $\Omega = \lfloor \frac{L}{\Lambda_1} \rfloor > \frac{L}{\Lambda_1} - 1$, we rewrite (A.1) and (A.2) as

$$\Lambda_1 \left( \frac{L}{\Lambda_1} - 1 \right) \left( \frac{L}{\Lambda_1} - 2 \right) \geq \left( \frac{\Lambda_2}{T_s} - 1 \right) \frac{\Lambda_2}{T_s} \tag{A.4}$$

$$\frac{T_s - k}{T_s}\Lambda_1 \left( \frac{L}{\Lambda_1} - 1 \right)^2 \geq \left( \frac{\Lambda_2}{T_s} \right)^2 \tag{A.5}$$

We carry out the minimization with respect to $(\Lambda_1, \Lambda_2)$ by first finding the two minimums with respect to the constraints (C1)/(C2) and (C1)/(C3), respectively, and then select the smaller one of these two. Using the simplified assumption [28] that the cardinalities of $\Lambda_1$ and $\Lambda_2$ are the same and to distinguish the two candidate minimums, we set $\Lambda_1 = \Lambda_2 = \Lambda_3$ in (A.5) and $\Lambda_1 = \Lambda_2 = \Lambda_4$ in (A.5) so that the above two inequalities become

$$\Lambda_3^3 - (T_s + 2T_s^2)\Lambda_3^2 + 3T_s^2\Lambda_3 L - T_s^2 L^2 \leq 0 \tag{A.6}$$

$$\Lambda_4^3 - T_s(T_s - k)\Lambda_4^2 + 2T_s(T_s - k)\Lambda_4 L - T_s(T_s - k)L^2 \le 0 \tag{A.7}$$

By defining $X_1 = \Lambda_3 - \frac{T_s - 2T_s^2}{3}$ and $X_2 = \Lambda_4 - \frac{T_s(T_s - k)}{3}$, we rewrite the above inequalities as

$$X_1^3 + \left(3T_s^2 L - \frac{1}{3}(T_s + 2T_s^2)^2\right)X_1 + \left(-T_s^2 L^2 + (T_s^3 + 2T_s^4)L - \frac{2}{27}(T_s + 2T_s^2)^3\right) \le 0 \tag{A.8}$$

$$X_2^3 + \left(2T_s(T_s - k)L - \frac{1}{3}T_s^2(T_s - k)^2\right)X_2 + \left(-T_s(T_s - k)L^2 + \frac{2}{3}T_s^2(T_s - k)^2 L - \frac{2}{27}T_s^3(T_s - k)^3\right) \le 0 \tag{A.9}$$

Following the standard procedure for solving a cubic equation [30], we define

$$\begin{aligned}
p_1 &= 3T_s^2 L - \frac{1}{3}(T_s + 2T_s^2)^2, \\
q_1 &= -T_s^2 L^2 + (T_s^3 + 2T_s^4)L - \frac{2}{27}(T_s + 2T_s^2)^3, \\
p_2 &= 2T_s(T_s - k)L - \frac{1}{3}T_s^2(T_s - k)^2, \\
q_2 &= -T_s(T_s - k)L^2 + \frac{2}{3}T_s^2(T_s - k)^2 L - \frac{2}{27}T_s^3(T_s - k)^3.
\end{aligned}$$

If $L > \frac{10}{3}T_s^3 + T_s^2 - \frac{T_s}{3}$, then

$$\begin{aligned}
p_1 &= 3T_s^2 L - \frac{1}{3}(T_s + 2T_s^2)^2 = T_s^2\left(3L - \frac{4}{3}T_s^2 - \frac{4}{3}T_s - \frac{1}{3}\right) \\
&> T_s^2\left(10T_s^3 + 3T_s^2 - 1 - \frac{4}{3}T_s^2 - \frac{4}{3}T_s - \frac{1}{3}\right) > 0 \tag{A.10}
\end{aligned}$$

$$\begin{aligned}
p_2 &= 2T_s(T_s - k)L - \frac{1}{3}T_s^2(T_s - k)^2 = 2T_s(T_s - k)\left(L - \frac{1}{6}T_s(T_s - k)\right) \\
&> 2T_s(T_s - k)\left(L - \frac{1}{6}T_s(T_s - 1)\right) > 0 \tag{A.11}
\end{aligned}$$

$$\begin{aligned}
q_1 &= -T_s^2 L^2 + (T_s^3 + 2T_s^4)L - \frac{2}{27}(T_s + 2T_s^2)^3 < -T_s^2 L^2 + (T_s^3 + 2T_s^4)L \\
&= -T_s^2 L(L - T_s - 2T_s^2) < -T_s^2 L\left(\frac{10}{3}T_s^3 + T_s^2 - \frac{T_s}{3} - T_s - 2T_s^2\right) < 0 \tag{A.12}
\end{aligned}$$

$$\begin{aligned}
q_2 &= -T_s(T_s - k)L^2 + \frac{2}{3}T_s^2(T_s - k)^2 L - \frac{2}{27}T_s^3(T_s - k)^3 < -T_s(T_s - k)L^2 + \frac{2}{3}T_s^2(T_s - k)^2 L \\
&< -T_s(T_s - k)L\left(L - \frac{2}{3}T_s^2\right) < -T_s(T_s - k)L\left(\frac{10}{3}T_s^3 + T_s^2 - \frac{T_s}{3} - \frac{2}{3}T_s^2\right) < 0 \tag{A.13}
\end{aligned}$$

$$\begin{aligned}
p_1 - p_2 &= 3T_s^2 L - \frac{1}{3}(T_s + 2T_s^2)^2 - 2T_s(T_s - k)L + \frac{1}{3}T_s^2(T_s - k)^2 \\
&> 3T_s^2 L - \frac{1}{3}(T_s + 2T_s^2)^2 - 2T_s^2 L + \frac{1}{3}T_s^2 = T_s^2 L - \frac{4}{3}T_s^2 - \frac{4}{3}T_s \\
&> T_s^2\left(\frac{10}{3}T_s^3 + T_s^2 - \frac{T_s}{3}\right) - \frac{4}{3}T_s^2 - \frac{4}{3}T_s > 0 \tag{A.14}
\end{aligned}$$

$$
\begin{aligned}
q_2 - q_1 &= -T_s(T_s-k)L^2 + \frac{2}{3}T_s^2(T_s-k)^2L - \frac{2}{27}T_s^3(T_s-k)^3 + T_s^2L^2 - (T_s^3+2T_s^4)L + \frac{2}{27}(T_s+2T_s^2)^3 \\
&> kT_sL^2 + \frac{2}{3}T_s^2(T_s-k)^2L - (T_s^3+2T_s^4)L > LT_s\left(L + \frac{2}{3}T_s^2 - (T_s^2+2T_s^3)\right) \\
&> LT_s\left(\frac{10}{3}T_s^3 + T_s^2 - \frac{T_s}{3} - (T_s^2+2T_s^3)\right) > 0.
\end{aligned}
\tag{A.15}
$$

These results imply $(\frac{p_1}{3})^3 + (\frac{q_1}{2})^2 > 0$, $(\frac{p_2}{3})^3 + (\frac{q_2}{2})^2 > 0$ and

$$
\Lambda_3 \leq \frac{T_s + 2T_s^2}{3} + \sqrt[3]{-\frac{q_1}{2} + \sqrt{(\frac{p_1}{3})^3 + (\frac{q_1}{2})^2}} + \sqrt[3]{-\frac{q_1}{2} - \sqrt{(\frac{p_1}{3})^3 + (\frac{q_1}{2})^2}} \stackrel{def}{=} C
\tag{A.16}
$$

$$
\Lambda_4 \leq \frac{T_s(T_s-k)}{3} + \sqrt[3]{-\frac{q_2}{2} + \sqrt{(\frac{p_2}{3})^3 + (\frac{q_2}{2})^2}} + \sqrt[3]{-\frac{q_2}{2} - \sqrt{(\frac{p_2}{3})^3 + (\frac{q_2}{2})^2}} \stackrel{def}{=} D.
\tag{A.17}
$$

It can be shown that

$$
G = \sqrt[3]{-\frac{q_1}{2} + \sqrt{(\frac{p_1}{3})^3 + (\frac{q_1}{2})^2}} + \sqrt[3]{-\frac{q_1}{2} - \sqrt{(\frac{p_1}{3})^3 + (\frac{q_1}{2})^2}} > 0
\tag{A.18}
$$

$$
H = \sqrt[3]{-\frac{q_2}{2} + \sqrt{(\frac{p_2}{3})^3 + (\frac{q_2}{2})^2}} + \sqrt[3]{-\frac{q_2}{2} - \sqrt{(\frac{p_2}{3})^3 + (\frac{q_2}{2})^2}} > 0
\tag{A.19}
$$

are zeros of $f(x) = x^3 + p_1x + q_1$ and $h(x) = x^3 + p_2x + q_2$, respectively. As both $f(x)$ and $g(x)$ are monotonically increasing functions and $f'(x) = 3x^2 + p_1 > g'(x) = 3x^2 + p_2 > 0, \forall x$, $f(x) < g(x), \forall x < \hat{x}$, where $\hat{x}$ is the single intersection point given by

$$
\hat{x} = \frac{q_2 - q_1}{p_1 - p_2} > 0.
$$

The fact that

$$
f(\hat{x}) = (\frac{q_2-q_1}{p_1-p_2})^3 + p_1\frac{q_2-q_1}{p_1-p_2} + q_1 = (\frac{q_2-q_1}{p_1-p_2})^3 + \frac{p_1q_2-p_2q_1}{p_1-p_2} > (\frac{q_2-q_1}{p_1-p_2})^3 + \frac{p_2q_2-p_2q_1}{p_1-p_2} > 0.
$$

implies that the only real zero of $f(x), G$, is larger than that of $g(x), H$, and thus $C > D$.

Substituting $\Lambda_i = C$ into (A.3), we obtain an upper-bound with a very complicated expression. To have an upper-bound with a simpler form, we notice that $||\Lambda_i||_{M=T_cT_s} = 0$ gives

$$
\max \Lambda_i = T_sT_c \left\lfloor \frac{C}{T_s \cdot T_c} \right\rfloor > C - T_sT_c
\tag{A.20}
$$

Hence a less tight upper-bound is given by

$$
\begin{aligned}
w_{4,min,ibp} &\leq 4 + 4\alpha\left(\min_{\Lambda_i}\left\{\left\lceil\frac{T_s \cdot L}{\Lambda_i}\right\rceil - 1\right\}\right) + 4\beta = 4 + 4\alpha\left(\left\lceil\frac{T_sL}{T_sT_c\left\lfloor\frac{C}{T_sT_c}\right\rfloor}\right\rceil - 1\right) + 4\beta \\
&\leq 4 + 4\alpha\left(\left\lceil\frac{T_sL}{C - T_sT_c}\right\rceil - 1\right) + 4\beta < 4 + 4\alpha \cdot \frac{T_sL}{C - T_sT_c} + 4\beta
\end{aligned}
\tag{A.21}
$$

The upper-bound of weight-4 input sequence in the Case (a) is twice the upper-bound of weight-2 input sequence shown in (51).

Note that

$$E \stackrel{def}{=} C - \sqrt{T_s L}\frac{T_s + 2T_s^2}{3} + \sqrt[3]{-\frac{q_1}{2} + \sqrt{(\frac{p_1}{3})^3 + (\frac{q_1}{2})^2}} + \sqrt[3]{-\frac{q_1}{2} - \sqrt{(\frac{p_1}{3})^3 + (\frac{q_1}{2})^2}} - \sqrt{T_s L} \quad \text{(A.22)}$$

and

$$\left(E - \frac{T_s + 2T_s^2}{3} + \sqrt{T_s L}\right)^3 = -q_1 - p_1\left(E - \frac{T_s + 2T_s^2}{3} + \sqrt{T_s L}\right). \quad \text{(A.23)}$$

In other words, $E$ is a zero of the polynomial

$$g(x) = \left(x - \frac{T_s + 2T_s^2}{3} + \sqrt{T_s L}\right)^3 + p_1\left(x - \frac{T_s + 2T_s^2}{3} + \sqrt{T_s L}\right) + q_1$$

which, like $f(x)$ defined before, is a monotonically increasing function and has only one real zero. For $T_s \geq 2$,

$$
\begin{aligned}
g(0) &= \left(-\frac{T_s + 2T_s^2}{3} + \sqrt{T_s L}\right)^3 + p_1\left(-\frac{T_s + 2T_s^2}{3} + \sqrt{T_s L}\right) + q_1 \\
&= -T_s^2 L^2 + \left(T_s^{\frac{3}{2}} + 3T_s^{\frac{5}{2}}\right)L^{\frac{3}{2}} - (T_s^2 + 2T_s^3)L \\
&< L^{\frac{3}{2}}\left(-T_s^2 L^{\frac{1}{2}} + \left(T_s^{\frac{3}{2}} + 3T_s^{\frac{5}{2}}\right)\right) \\
&< L^{\frac{3}{2}}\left(-T_s^2\left(\frac{10}{3}T_s^3 + T_s^2 - \frac{T_s}{3}\right)^{\frac{1}{2}} + \left(T_s^{\frac{3}{2}} + 3T_s^{\frac{5}{2}}\right)\right) < 0
\end{aligned}
\quad \text{(A.24)}
$$

The last inequality holds because both $T_s^2\left(\frac{10}{3}T_s^3 + T_s^2 - \frac{T_s}{3}\right)$ and $T_s^{\frac{3}{2}} + 3T_s^{\frac{5}{2}}$ are positive real numbers and

$$
\begin{aligned}
T_s^4\left(\frac{10}{3}T_s^3 + T_s^2 - \frac{T_s}{3}\right) - (T_s^{\frac{3}{2}} + 3T_s^{\frac{5}{2}})^2 &= \frac{10}{3}T_s^7 + T_s^6 - \frac{T_s^5}{3} - 9T_s^5 - 6T_s^4 - T_s^3 \\
&> \frac{40}{3}T_s^5 + 2T_s^5 - \frac{T_s^5}{3} - 9T_s^5 - 6T_s^4 - T_s^3 \\
&> 12T_s^5 - 6T_s^4 - T_s^3 > 0.
\end{aligned}
$$

Hence $E$ is positive and so

$$4 + 4\alpha \cdot \frac{T_s L}{C - T_s T_c} + 4\beta < 4 + 4\alpha\frac{T_s L}{\sqrt{T_s L} - T_s T_c} + 4\beta = 2\left(2 + 2\alpha\frac{T_s L}{\sqrt{T_s L} - T_s T_c} + 2\beta\right). \quad \text{(A.25)}$$

TABLE I

IBP ALGORITHM

<u>Variables</u>

L-block length

N-total number of blocks

K-block number index

D(m,k)-data on the $k$th block $m$th position

<u>Recursion</u>

for K=0 to N-1

    for i=0 to i=S-1

      if (K-i > 0)

        if (K mod (2·(i+1)) < i+1)

          set m=2·i+1

        else

          set m=2·i+2

        end

        while (m < L)

          swap D(m,K) and D(m,K-i-1)

          set m=m+2S+1

        end

      end

    end

end

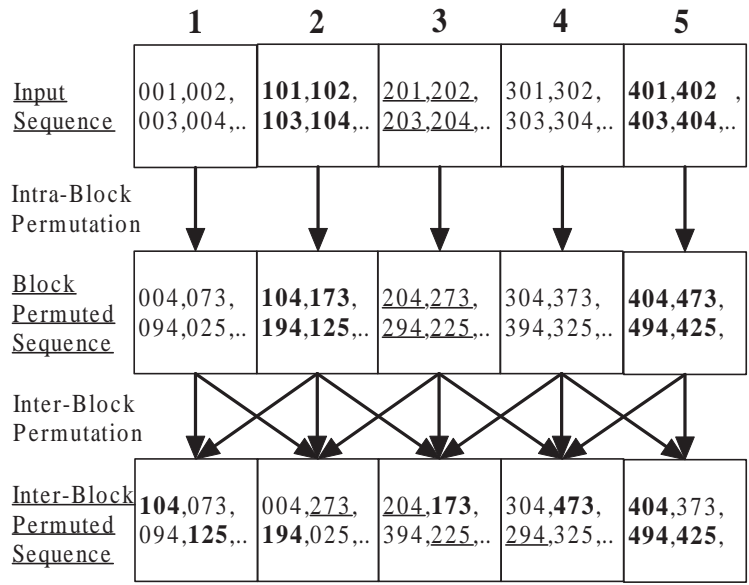

Fig. 1. An inter-block permutation interleaving procedure.

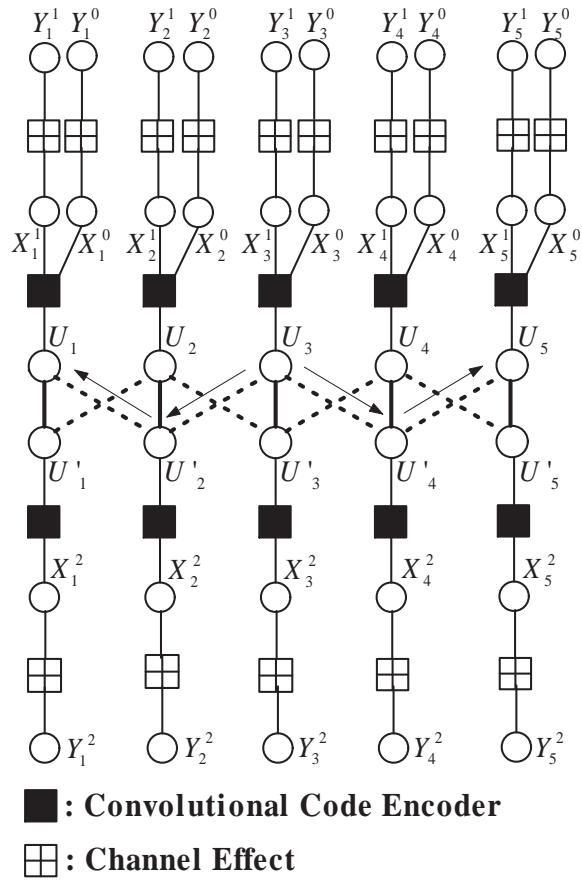

**■ : Convolutional Code Encoder**

**⊞ : Channel Effect**

Fig. 2. Factor graph representation and information flow of an IBPTC and a classic TC.

Classic Turbo Code

| APP Decoding Round \ Block Index | 1 | 2 | 3 | 4 | 5 | 6 | 7 |
|---|---|---|---|---|---|---|---|
| 1 | 1 | 5 | 9 | 13 | 17 | 21 | 25 |
| 2 | 2 | 6 | 10 | 14 | 18 | 22 | 26 |
| 3 | 3 | 7 | 11 | 15 | 19 | 23 | 27 |
| 4 | 4 | 8 | 12 | 16 | 20 | 24 | 28 |

IBPTC

| APP Decoding Round \ Block Index | 1 | 2 | 3 | 4 | 5 | 6 | 7 |
|---|---|---|---|---|---|---|---|
| 1 | 1 | 2 | 4 | 7 | 11 | 15 | 19 |
| 2 | 3 | 5 | 8 | 12 | 16 | 20 | 23 |
| 3 | 6 | 9 | 13 | 17 | 21 | 24 | 26 |
| 4 | 10 | 14 | 18 | 22 | 25 | 27 | 28 |

Fig. 3. A comparison of exemplary decoding schedules for classic TC and IBPTC when decoding 7 blocks with 2 iterations (four decoding rounds). The numbers in the two rectangular grid-like tables represent the order the APP decoder performs decoding. Hence the first block of the classic TC is decoded by the first 4 decoding rounds (the leftmost column) but that of the IBPTC is decoded by the first, third, sixth and tenth decoding rounds; see Section III for detailed discussion.

| APP Decoding Round \ Block Index | 1 | 2 | 3 | 4 | 5 | 6 | 7 |
|---|---|---|---|---|---|---|---|
| 1 | $\mathbf{a}_{11}$ | $\mathbf{b}_{11}$ | $\mathbf{c}_{11}$ | $\mathbf{d}_{11}$ | $\mathbf{a}_{21}$ | $\mathbf{b}_{21}$ | $\mathbf{c}_{21}$ |
| 2 | $\mathbf{b}_{12}$ | $\mathbf{c}_{12}$ | $\mathbf{d}_{12}$ | $\mathbf{a}_{22}$ | $\mathbf{b}_{22}$ | $\mathbf{c}_{22}$ | $\mathbf{d}_{22}$ |
| 3 | $\mathbf{c}_{13}$ | $\mathbf{d}_{13}$ | $\mathbf{a}_{23}$ | $\mathbf{b}_{23}$ | $\mathbf{c}_{23}$ | $\mathbf{d}_{23}$ | $\mathbf{a}_{33}$ |
| 4 | $\mathbf{d}_{14}$ | $\mathbf{a}_{24}$ | $\mathbf{b}_{24}$ | $\mathbf{c}_{24}$ | $\mathbf{d}_{24}$ | $\mathbf{a}_{34}$ | $\mathbf{b}_{34}$ |

Fig. 4. Four-processor decoding schedule for an IBPTC where $z_{ij}$ indicates that the APP decoder $z$ is performing the $j$th decoding round of its $i$th decoding phase; each arrowed dashed slant line from upper-right to lower-left represents a decoding phase for a certain APP decoder.

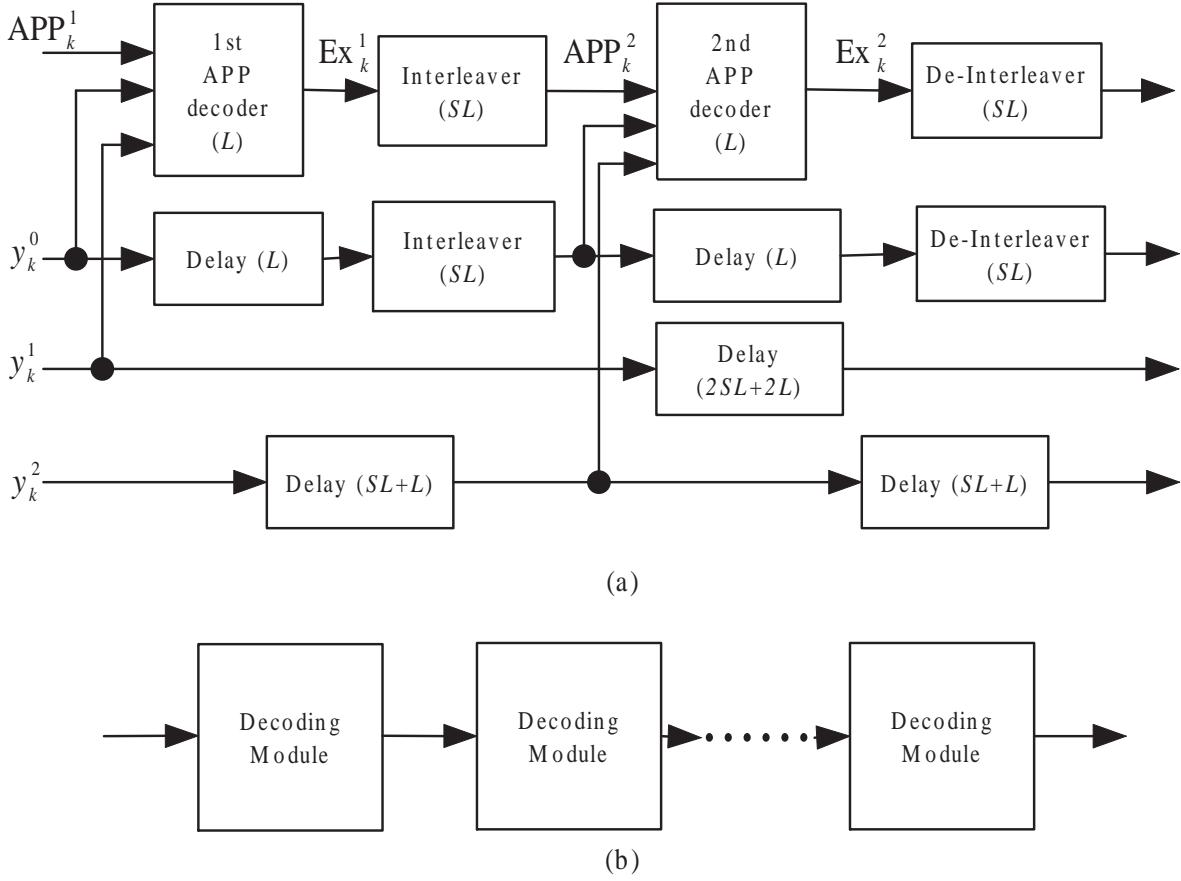

(a)

(b)

Fig. 5. (a) An IBPTC decoding module for 1 iteration; (b) An IBPTC pipeline decoder. $y_k^i$ is the received sample corresponding to the transmitted bits $x_k^i$; $\text{APP}_k^1$ and $\text{Ex}_k^1$ represent the a priori and the extrinsic information associated with the $k$th information bit $u_k$; $\text{APP}_k^2$ and $\text{Ex}_k^2$ represent the a priori and the extrinsic information associated with the $k$th interleaved information bit $u_k'$.

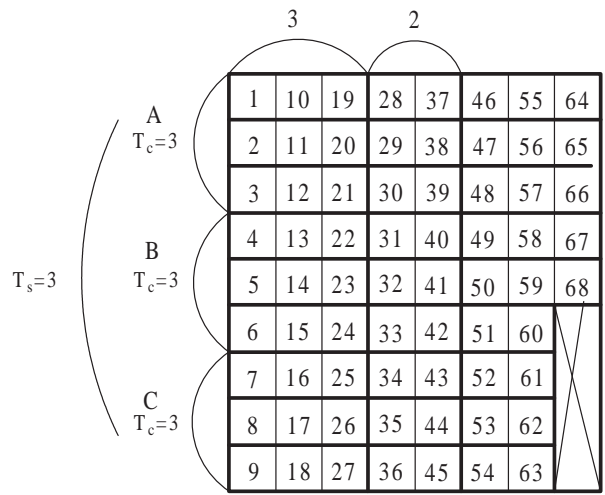

Fig. 6. Partition of equivalence classes into subsets and IBP interleaving; $L = 68$, $\Lambda = 27$, $T_c = T_s = 3$.

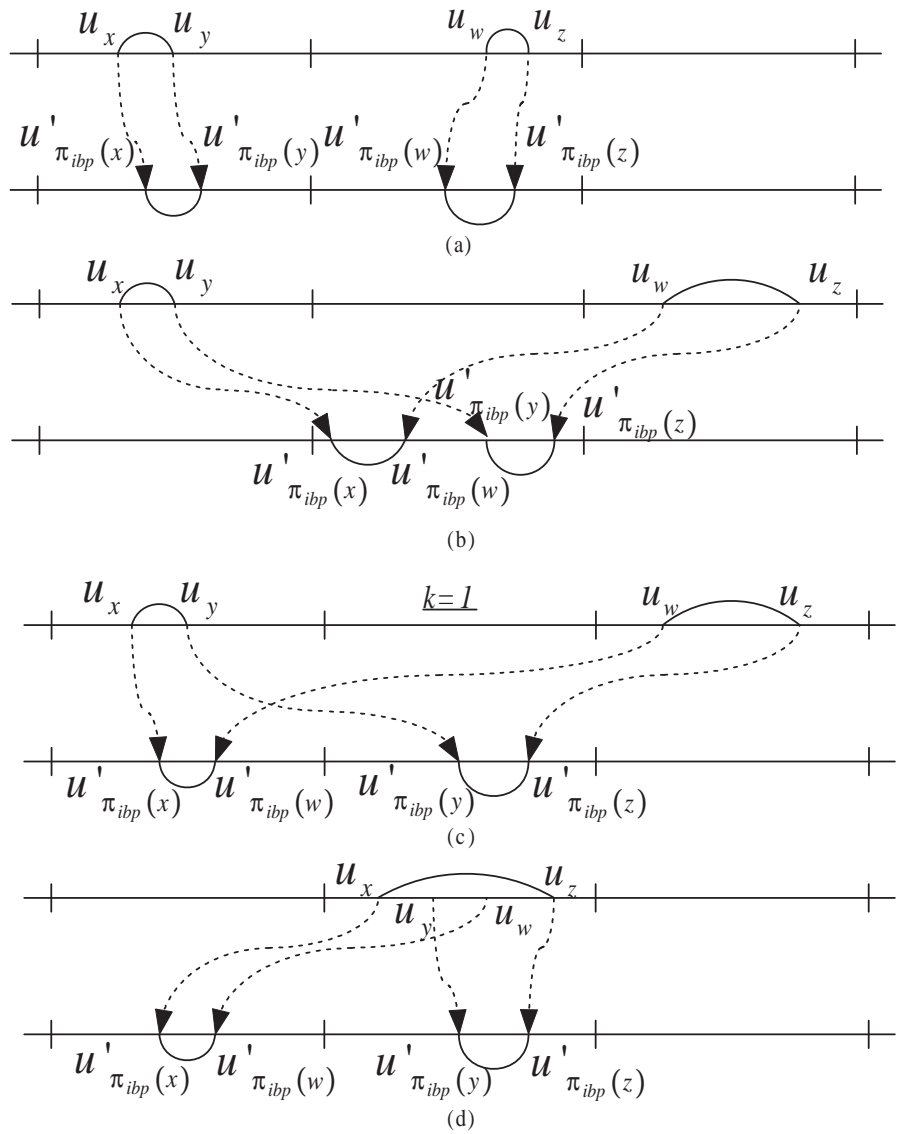

Fig. 7. Pre- and post-interleaving nonzero coordinate distributions of weight-4 input sequences that result in low-weight IBPTC codewords.

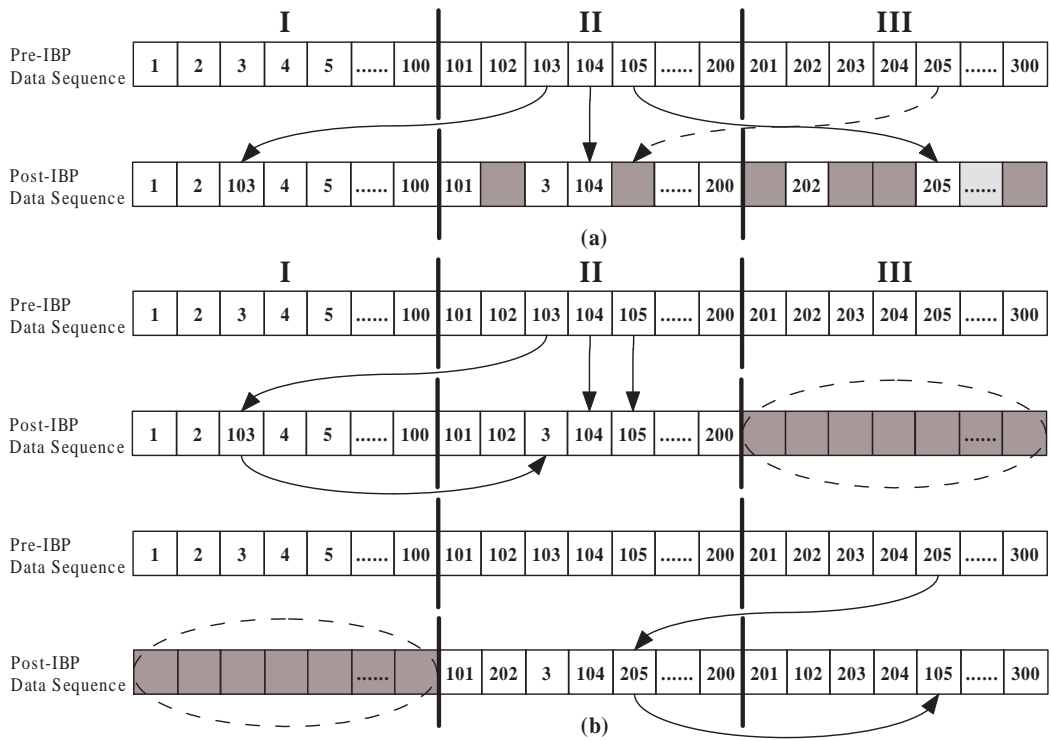

Fig. 8. (a) Inter-block interleaving using a non-swap structure ($S = 1$); (b) inter-block interleaving using the swap structure. When one starts to interleave (or de-interleave) Block III, Block I has been completely interleaved (or de-interleaved), its content was dumped and the corresponding space is emptied and becomes available for storing new content again. The storage spaces enclosed by dotted ellipses are thus not needed.

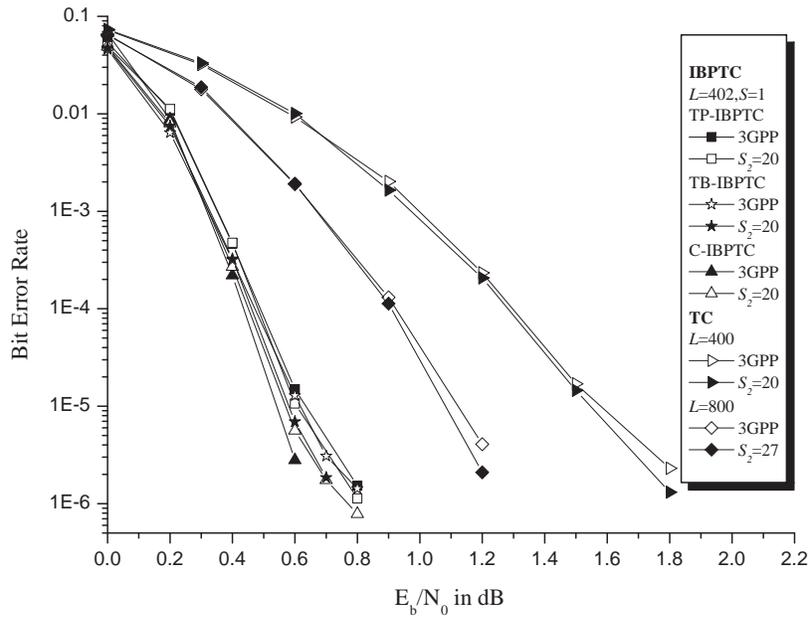

Fig. 9. BER performance of IBPTCs with SRID $\approx 800$, block size $L = 402$ and interleaver span $S = 1$. For comparison purpose, performance of the classic TC with $L = 400, 800$ are also given.

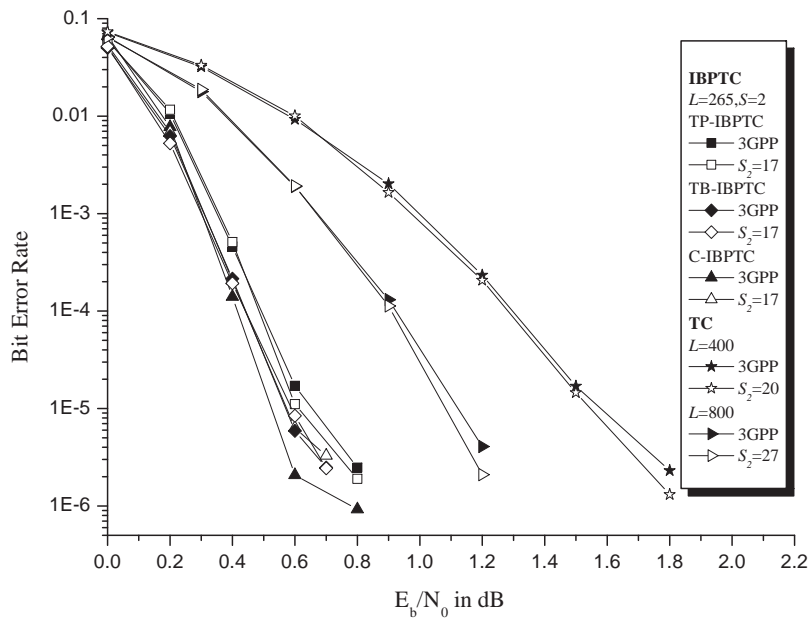

Fig. 10. BER performance of IBPTCs with SRID $\approx 800$, block size $L = 265$ and interleaver span $S = 2$. For comparison purpose, performance of the classic TC with $L = 400, 800$ are also given.

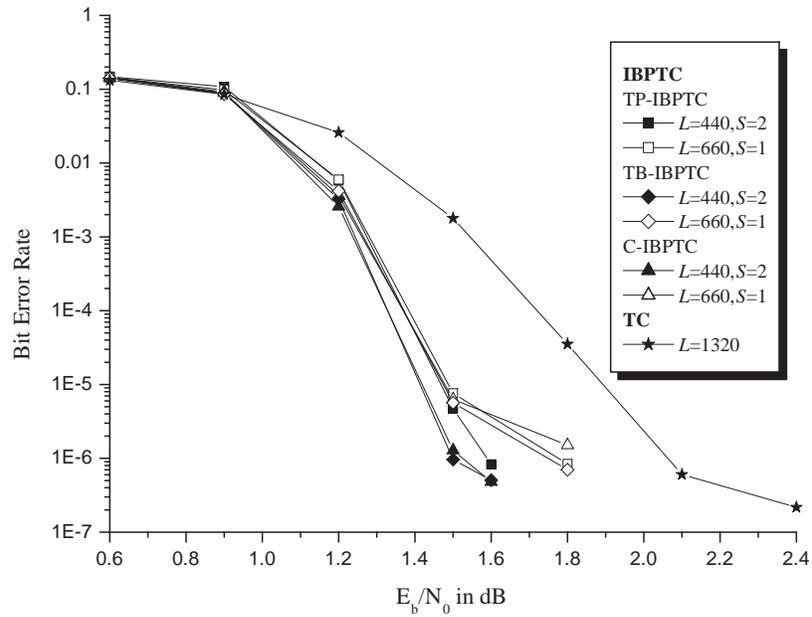

Fig. 11. BER performance of IBPTCs and the classic TC with SRID =1320 and the 3GPP interleaver.

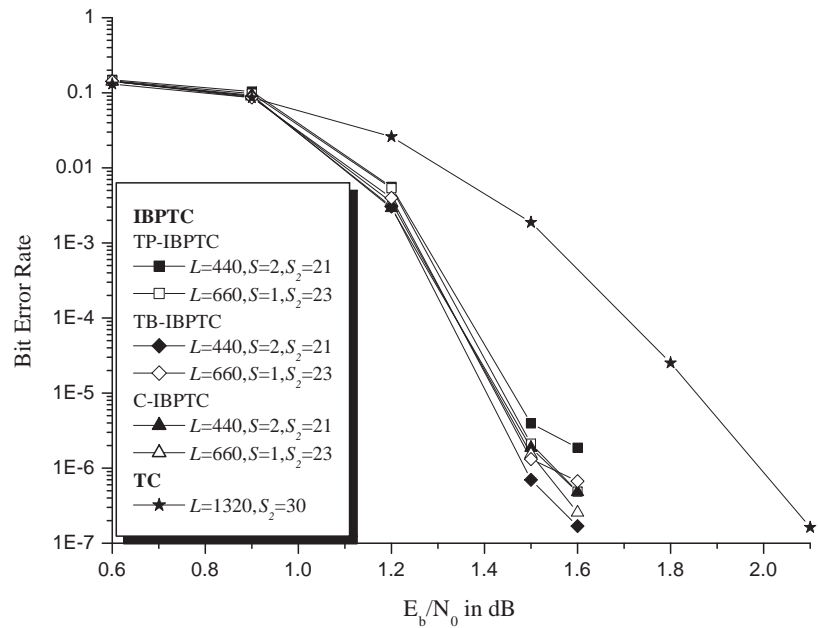

Fig. 12. BER performance of IBPTCs and the classic TC with SRID = 1320 and the modified semi-random interleaver.

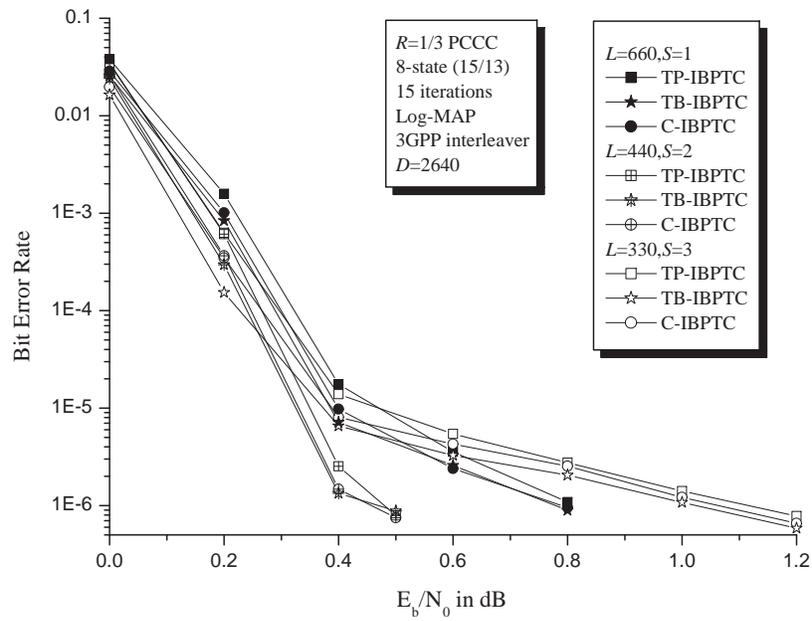

Fig. 13. Influence of the interleaver span on the BER performance for various IBPTCs with SRID =1320.

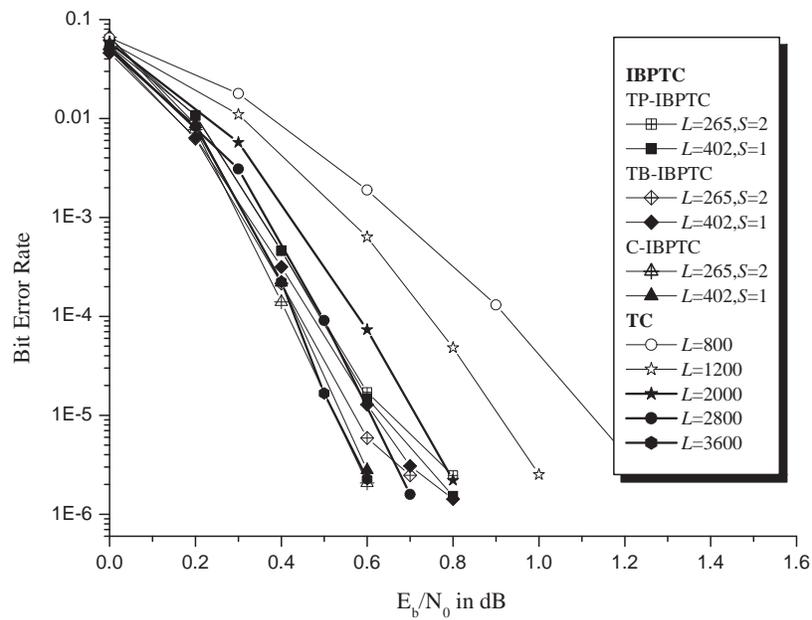

Fig. 14. BER comparison of IBPTCs and the 3GPP defined turbo code of various block sizes.

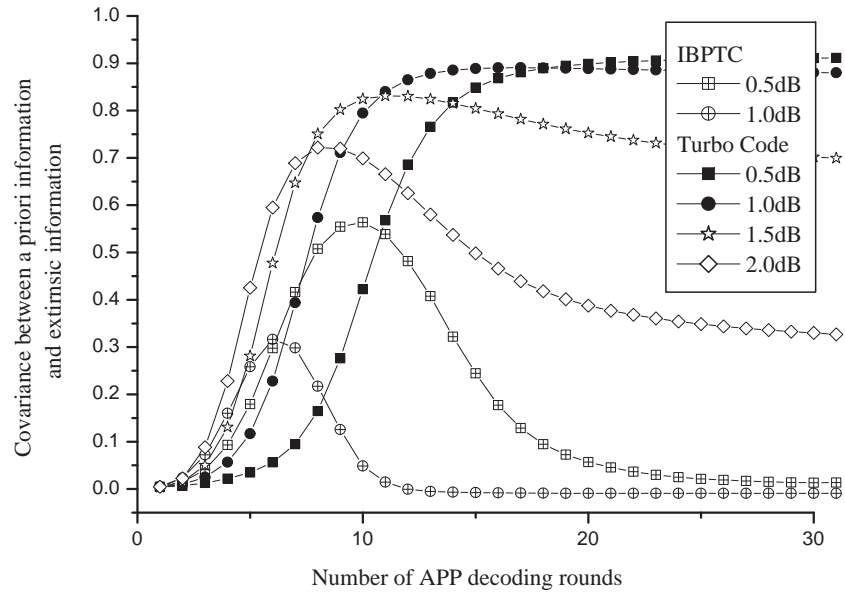

Fig. 15. Covariance between a priori information input and extrinsic information output.

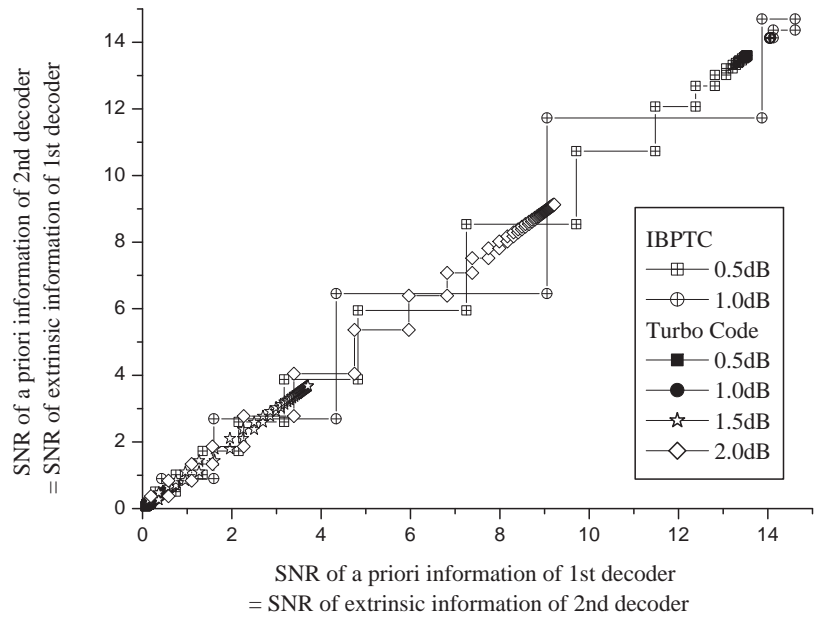

Fig. 16. SNR evolution chart behavior of the IBPTC and the classic TC at different $E_b/N_0$'s.

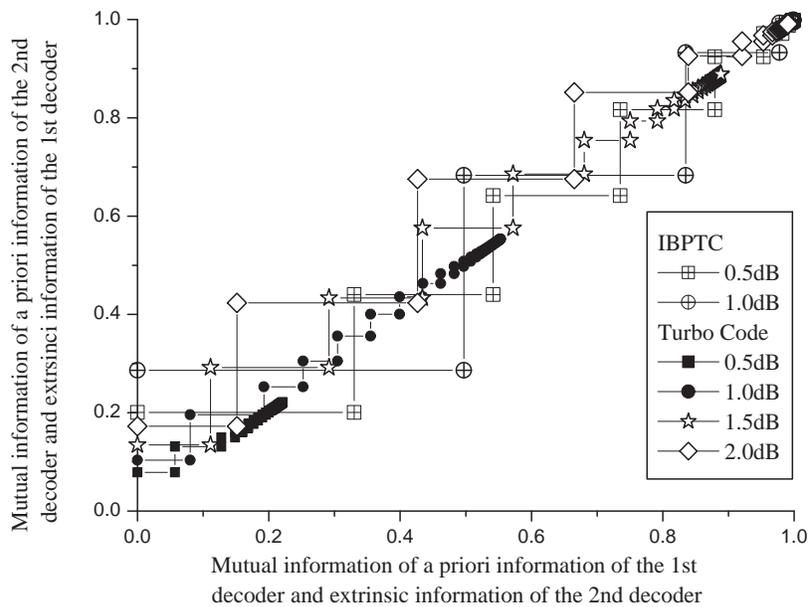

Fig. 17. Exit chart performance of the IBPTC and the classic TC at different $E_b/N_0$'s.